\documentclass[sigconf]{acmart}
\AtBeginDocument{ \providecommand\BibTeX{{ \normalfont B\kern-0.5em{\scshape i\kern-0.25em b}\kern-0.8em\TeX}}}

\usepackage{algorithmic}
\usepackage{amsmath}
\usepackage{amsmath,amsfonts}
\usepackage{array}
\usepackage{booktabs}
\usepackage{caption}
\usepackage{comment}
\usepackage{enumerate,enumitem}
\usepackage{filecontents}
\usepackage{graphicx}
\usepackage{hyperref}
\usepackage{listings}
\usepackage{soul}
\usepackage{stfloats}
\usepackage{subcaption}
\usepackage{textcomp}
\usepackage{tikz}
\usepackage{url}
\usepackage{verbatim}
\usepackage{xcolor,flushend}
\usepackage{xspace,xcolor,yfonts,multirow}

\newcommand{\Pome}{{Pome}\xspace}
\newcommand{\SST}{{SSTable}\xspace}
\newcommand{\SSTs}{{SSTables}\xspace}

\newcommand*\mininumbercircled[1]{\tikz[baseline=(char.base)]{
\node[fill=white, text=black, shape=circle,draw,inner sep=0.1pt] (char) {#1};}}

\copyrightyear{2026}
\acmYear{2026}
\setcopyright{cc}
\setcctype{by-nc-nd}
\acmConference[HPDC '26]{The 35th International Symposium on High-Performance Parallel and Distributed Computing}{July 13--16, 2026}{Cleveland, OH, USA}
\acmBooktitle{The 35th International Symposium on High-Performance Parallel and Distributed Computing (HPDC '26), July 13--16, 2026, Cleveland, OH, USA}
\acmDOI{10.1145/3806645.3807603}
\acmISBN{979-8-4007-2640-8/2026/07}

\begin{document}

\title{\Pome: Parallelizing I/Os and Computations for Efficient LSM-tree-based Data Storage}

\author{Yanpeng Hu$^*$}
\affiliation{	\institution{ShanghaiTech University}
\city{Shanghai}
\country{China} }

\author{Li Zhu}
\authornote{Y. Hu and L. Zhu contributed equally to this work.}
\affiliation{	\institution{ShanghaiTech University}
\city{Shanghai}
\country{China} }

\author{Lei Jia}
\affiliation{	\institution{ShanghaiTech University}
\city{Shanghai}
\country{China} }

\author{Chundong Wang}
\authornote{C. Wang is the corresponding author (cd\_wang@outlook.com).}
\affiliation{	\institution{ShanghaiTech University}
\city{Shanghai}
\country{China} }

\renewcommand{\shortauthors}{Y. Hu, L. Zhu, L. Jia, and C. Wang}
\begin{abstract} CPU computations and I/O operations are fundamental to data storage systems. Storage systems conduct computations with their user threads, such as sorting data for orderliness. They handle I/Os through system calls (syscalls) including file write, read, and {\tt fsync}, which the OS's kernel threads perform with storage devices.

    Today, LSM-tree-based storage systems are widely deployed in production environments. Compaction is an essential operation that LSM-tree employs to maintain its tiered tree-like structure by re-sorting and re-storing data through computations and I/Os, respectively. In this paper, we first overhaul the procedure of a compaction. We find that computations and I/Os execute in a sequential order. After re-sorting data, the user thread waits for a kernel thread to complete file write and {\tt fsync} I/Os. These costly synchronous I/Os create a severely long critical path that affects the performance of LSM-tree. To address this issue, we propose \underline{\textbf{p}}arallelizing I/\underline{\bf O}s and co\underline{\bf m}putations for efficient LSM-tree-based data storag\underline{\bf e} (\textbf{Pome}). Pome decouples computations from I/Os within each compaction by referring to its new protocol that moves I/O operations out of the critical path. To this end, it conducts asynchronous I/Os by using io\_uring. Furthermore, regarding the potential I/O congestion caused by accelerated compactions,
    \Pome incorporates an adaptive I/O rate limiter to achieve smooth execution. We prototype Pome on top of RocksDB. Experimental results demonstrate that Pome significantly improves the performance of RocksDB and outperforms several state-of-the-art LSM-tree variants.
\end{abstract}



\maketitle

\section{Introduction}\label{sec:introduction}

I/Os and computations are fundamental for data storage. I/Os are used to load and store on-disk data through system calls (syscalls), such as file write, read, and {\tt fsync}. Computations include sorting data for orderly organization, compressing and decompressing to save space, encrypting/decrypting data for security, and so on. Computations and I/Os are typically handled by user and kernel threads, respectively. In today's production environments, the structure of log-structured merge-tree (LSM-tree) is widely used to build key-value (KV) store systems~\cite{LSM:LevelDB,LSM:RocksDB,LSM:in-SSD-index:OSDI-2021,TriangleKV:LDS:TPDS-2022,LSM:FPGA-accelerated-compaction:FAST-2020,LSM:SpanDB:FAST-2021,10.1145/3588195.3595949,LSM:ELECT:FAST-2024}. In this paper, we mainly take LSM-tree to exploit the potential of parallelizing computations and I/Os for efficient data storage.

RocksDB is a representative LSM-tree-based KV store~\cite{LSM:RocksDB}. Without loss of generality, we adopt the terminologies and strategies of RocksDB to explain how a LSM-tree operates. LSM-tree inserts KV pairs that clients submit into an in-memory structure called {\em memtable}. Once a memtable is full, it is transformed and {\em flush}ed into a {\it sorted string table} (\SST) file. LSM-tree manages \SSTs across multiple on-disk {\em levels}. A freshly flushed \SST\ is persisted and placed at the top level, i.e., $L_0$. LSM-tree defines that the capacity limit of level $L_{i+1}$ is ten times that of $L_i$ ($i\ge0$). When $L_i$ is full, LSM-tree initiates a {\em compaction} job with a user thread. In short, each compaction involves 1) merge-sorting KV pairs in \SSTs selected from levels $L_i$ and $L_{i+1}$, 2) generating new $L_{i+1}$ \SSTs and filling in them using sorted KV pairs through file write syscalls, and 3) ensuring the persistence of new $L_{i+1}$ \SSTs through {\tt fsync} syscalls. The user thread concludes the compaction upon receiving a completion or failure signal for {\tt fsync}s from a kernel thread.

More importantly, in the process of compaction, computations and I/Os follow a {\em sequential} execution order. As {\tt fsync} is time-consuming \cite{FS:BarrierFS:FAST-2018,FS:FastCommit:ATC-2024}, the user thread of LSM-tree has to stall until the kernel thread responds. During this time, the user thread stays idle, merely waiting, leading to a severely long critical path. In terms of hardware utilization, CPU and storage device alternately turn to be underutilized under this sequential execution model.

LSM-tree-based KV stores, including RocksDB, have considered parallelizing compactions at the job level. They support setting up multiple user threads to proceed concurrent compaction jobs. However, our study indicates that while the performance of RocksDB initially increases with the addition of a few more threads, it soon plateaus with furthermore compaction threads. This stagnation occurs because the user thread for each compaction job still has to wait for its corresponding kernel thread. Moreover, excessive {\tt fsync}s, which many kernel threads simultaneously push on, lead to contention and substantially increase the waiting time~\cite{FS:BarrierFS:FAST-2018,FS:FastCommit:ATC-2024,FS:Sync+Sync:Security-2024}.

We aim to reshape the execution model of a compaction job for LSM-tree, and deeply analyze the process and purpose of compaction. In a compaction, the user thread firstly conducts computations (merge-sort) and then waits for a kernel thread that mainly handles {\tt fsync}s. These costly {\tt fsync}s I/Os are expended on persisting new SSTable files to ensure durability. Whereas, it is important to note that compaction does not generate new data. Instead, KV pairs are loaded from input SSTables, while the compaction just sorts and re-stores them into output SSTables. If the input SSTables are already durably stored, there is no urgent need to persist the output SSTables immediately. Thus, the {\em existing durability} mitigates the pressing requirement for kernel threads to perform synchronous {\tt fsync}s on the critical path. As a result, we posit that by offloading all or some of heavy I/O operations from the critical path, the performance of LSM-tree shall gain a dramatic improvement.

This inspiring observation motivates us to parallelize computations and I/Os in each compaction. However, several challenges arise. First, the existing durability of SSTables involves dependency: a compaction's output \SSTs serve as inputs for future compactions. Keeping all old durable \SSTs to back up new ones is impractical due to high spatial cost. We therefore need to design a sound protocol to efficiently transition durability from input to output \SSTs and remove the former at an appropriate time. 

Second, durability enforcement for output \SSTs remains necessary but should occur outside of the critical path. Asynchronous I/O supports such as Linux AIO~\cite{aio:linux-man}, ASIO~\cite{Asio-lib}, and io\_uring~\cite{iouring} enable such decoupling. Among them, io\_uring stands out because it is efficient, flexible, and not bound to specific I/O modes or particular file systems \cite{10.1145/3731545.3735116}. We leverage io\_uring as a tool to conduct asynchronous write I/Os, allowing the user thread to proceed to the next compaction while the kernel thread persists data for the previous one. This effectively overlaps the I/Os of the current compaction with the computations of the next, achieving parallelization.

Third, by not synchronously waiting for the kernel thread, a user thread can finish its compaction job faster and quickly initiate subsequent compactions, issuing more asynchronous I/Os. However, excessive asynchronous I/Os may congest the storage device and degrade performance. We thus require a rate-limiting mechanism to regulate compaction pace, preventing I/O congestion and enabling smooth execution.

Accordingly, we propose {\bf \Pome}, a novel algorithm to \underline{\textbf{p}}arallelize I/\underline{\bf O}s and co\underline{\bf m}putations for efficient LSM-tree-based data storag\underline{\bf e}.
\Pome dedicates user threads to computations and prevents them from stalling on kernel threads that handle synchronous I/Os. This substantially shortens the critical path of serving client requests. The key aspects of \Pome, as well as the main contributions of this paper, are summarized as follows.

\begin{itemize}    \item We dissect the   compaction procedure of LSM-tree and reveal that existing data durability allows us to decouple CPU computations from file I/Os in a compaction.     \item We parallelize user and kernel threads by overlapping file I/Os of one compaction with CPU computations of the next, effectively shortening the critical path and improving the utilization of both CPU and storage device.
    \item We introduce a new compaction protocol to guarantee both durability and accessibility of data under asynchronous file write and {\tt fsync} operations, without undermining the rationality and correctness of LSM-tree-based data storage.
    \item We incorporate an adaptive I/O rate limiter to prevent excessive asynchronous I/Os from causing congestion, thereby preserving smooth system execution.
\end{itemize}

We prototype\footnote{The source code of \Pome is available at \href{https://github.com/toast-lab/LSM-Pome}{\url{https://github.com/toast-lab/LSM-Pome}}.} and evaluate \Pome with extensive experiments.
\Pome dramatically boosts RocksDB's performance. For example, when serving typical write-intensive workloads, \Pome shortens the client-facing 99th percentile (99P) tail latency by up to 3.0$\times$ compared to RocksDB, and also significantly outperforms state-of-the-art (SoTA) LSM-tree variants.

The remainder of this paper is organized as follows. Section~\ref{sec:background} presents the background of LSM-tree and asynchronous I/O. Section~\ref{sec:motivation} describes our motivational study. Section~\ref{sec:design} details the design and implementation of \Pome. Section~\ref{sec:evaluation} evaluates \Pome, and Section~\ref{sec:relatedWork} compares it with related works. Section~\ref{sec:conclusion} concludes the paper.

\begin{figure}[t]
    \begin{center}
        \scalebox{1.0}{\includegraphics[width=\columnwidth]{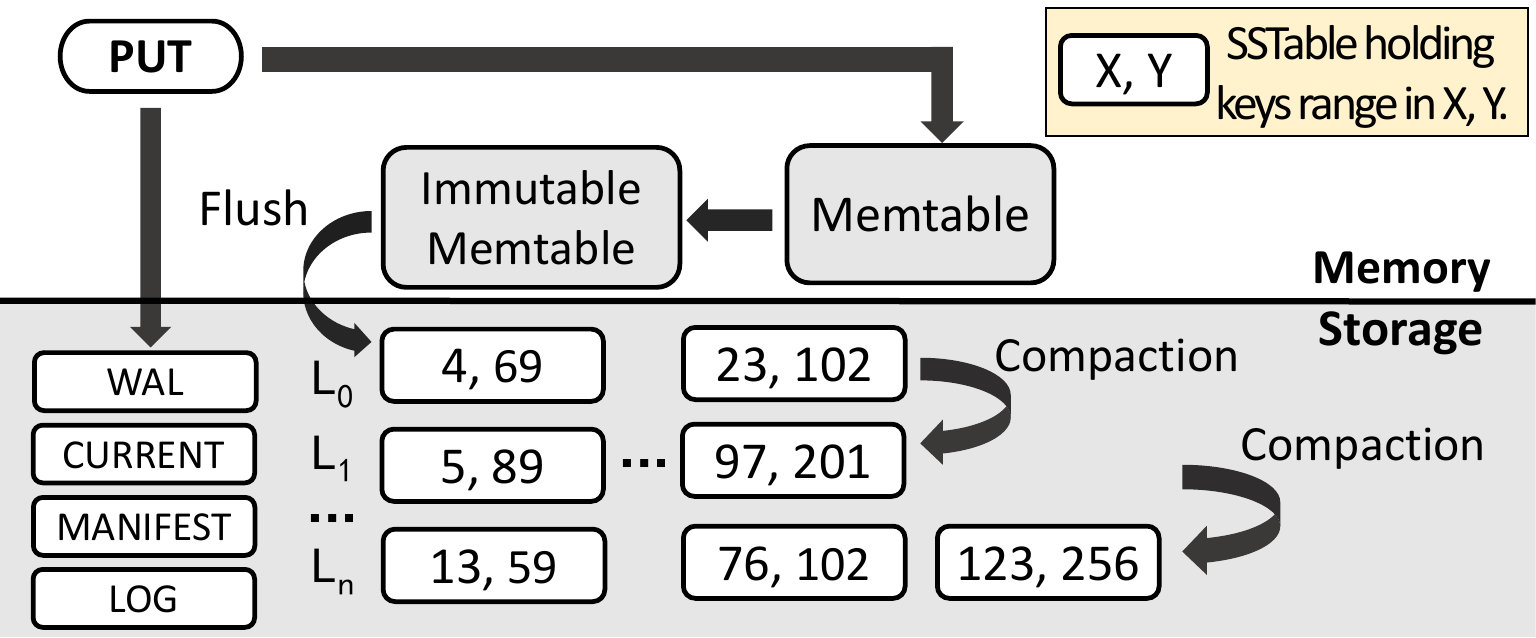}}
        \caption{The architecture of RocksDB}\label{fig:bg:Put}
    \end{center}
\end{figure}

\section{Background}\label{sec:background}

\textbf{I/Os and Computations.} File I/Os are fundamental to data storage systems that rely on syscalls such as file read, write and \texttt{fsync} to load and store data. A kernel thread executes one such syscall and returns a completion or failure signal, while a user thread of the storage system synchronously waits. Given the high cost of synchronous file operations, especially \texttt{fsync} or its variant \texttt{fdatasync}~\cite{FS:FastCommit:ATC-2024,FS:BarrierFS:FAST-2018}, the waiting time can be severely long.

Computations are also essential for data storage. For example, (re-)sorting data is frequently performed in SQL databases and NoSQL KV stores to maintain the orderliness for efficient indexing. Some storage systems compress or deduplicate data to save disk space, while others perform encryption and decryption for data security. More importantly, data sequentially undergoes CPU computations and I/Os in these systems. In short, data is persistently stored after being sorted, compressed, deduplicated, or encrypted.

\textbf{LSM-tree and RocksDB.} The {\em sequential} execution model between computations and I/Os generally entails a long critical path. In this paper, we take LSM-tree as a representative to illustrate that such a sequential model leads to severe performance penalty. However, we are able to reshape it into a new \textit{parallel} execution model and accordingly gain high efficiency.

RocksDB is a typical KV store built on the concept of LSM-tree.
\autoref{fig:bg:Put} shows an example of it, which consists of in-memory and on-disk components, resembling a tiered structure. RocksDB employs a skiplist ordered by keys as the in-memory \textit{memtable}. On receiving a KV pair in a \texttt{Put} request, RocksDB first appends the KV pair to a write-ahead log (WAL), then inserts it into the memtable. When the memtable reaches a predefined size limit (64\,MB by default), RocksDB renders it immutable, to be used for serving lookups only. Next, RocksDB creates a new mutable memtable. A user thread then transforms and flushes\footnote{ In this paper, we use the terminologies of RocksDB for presentation, such as {\em flush} and {\em compaction}. Researchers may use `flush' to 	describe a program calling {\tt fsync} to store a file, 		which we refer to as `persist' 	to distinguish from the {\em flush} of RocksDB.} the immutable memtable to an \textit{SSTable} file, where keys remain sorted to preserve orderliness.

RocksDB places each flushed SSTable on the top on-disk level, denoted as $L_0$, and ensures its persistence via \texttt{fsync} syscall. The corresponding WAL file is then safely deleted. RocksDB organizes on-disk data across multiple levels, each with an exponentially larger capacity bound than the previous one: by default, the capacity of level $L_{i+1}$ is ten times that of level $L_i$ ($i \ge 0$). To maintain the level balance and control space amplification, RocksDB employs one or more user threads to perform \textit{compaction} jobs. Based on a scoring mechanism that reflects the fullness of each level, RocksDB selects the level $L_i$ whose capacity the most exceeds its corresponding bound as the source for compaction. During compaction, a user thread first identifies a set of SSTable files from $L_i$ and $L_{i+1}$ that have overlapped key ranges. It then loads and merge-sorts all involved KV pairs, writes the sorted results into new SSTable files at $L_{i+1}$, and synchronously persists them using \texttt{fsync}. Afterward, the input SSTable files are deleted, and the new ones become accessible as the valid data of level $L_{i+1}$. {\em Each compaction job thus follows a strictly {sequential} execution order as aforementioned, i.e., merge-sort computations, synchronous file writes and \texttt{fsync} operations}.

\textbf{Asynchronous I/Os.} Computations are executed on CPU cores, whereas I/O operations are performed by peripheral storage devices, e.g., NVMe SSD, outside of the CPU core. This physical separation implies an opportunity for I/Os to be issued and completed asynchronously. The Linux kernel has long provided native support for \textit{asynchronous I/O} (AIO), but its practical use is limited due to several inherent drawbacks. For instance, AIO is only available in the direct I/O mode and may exhibit non-deterministic blocking behavior under certain conditions~\cite{iouring}.

To address these issues, J. Axboe recently introduced the {io\_uring} framework to substitute AIO~\cite{iouring,AnIntrod6:online}. The {io\_uring} offers low-latency, feature-rich interfaces for asynchronous operations, while maintaining full kernel-space execution. For example, RocksDB developers have utilized it to accelerate {\tt MultiGet} that loads (reads) scattered KV pairs while HPC researchers leverage io\_uring for efficient checkpoint restoration~\cite{10.1145/3731545.3735116}. Unlike SPDK that is a user-space library implementing its own file system and driver stack for applications~\cite{LSM:SpanDB:FAST-2021,IO:XRP:OSDI-2022,10.1145/3731569.3764804}, {io\_uring} allows applications to run directly on mature kernel file systems (e.g., Ext4 or XFS) in either buffered or direct I/O mode. Programmers seeking asynchronous behavior may also use user-space frameworks such as the ASIO library for C++~\cite{Asio-lib}, or exploit the implicit asynchronous commit mechanism provided by Ext4 journaling~\cite{LSM:NobLSM:DAC-2022}. However, compared with these alternatives, {io\_uring} delivers higher flexibility, kernel-level stability, and broad compatibility, without any restriction to a particular programming language or file system. Therefore, we will take io\_uring as a supportive library in developing our design.

\section{Motivation}\label{sec:motivation}

We conduct a study to analyze the impact of sequentially executing computations and I/Os during compaction on the performance of LSM-tree. We set up RocksDB on our experimental platform and perform a series of tests. More details of the setup can be found in Section~\ref{sec:evaluation}. It is worth noting that many LSM-tree-based KV stores adopt similar compaction strategies like those of RocksDB, thereby exhibiting comparable behaviors and issues revealed in our study. Through this quantitative study, we gain following observations.

\begin{figure}[t]
    \begin{center}
        \scalebox{1.0}{\includegraphics[width=\columnwidth, page=1]{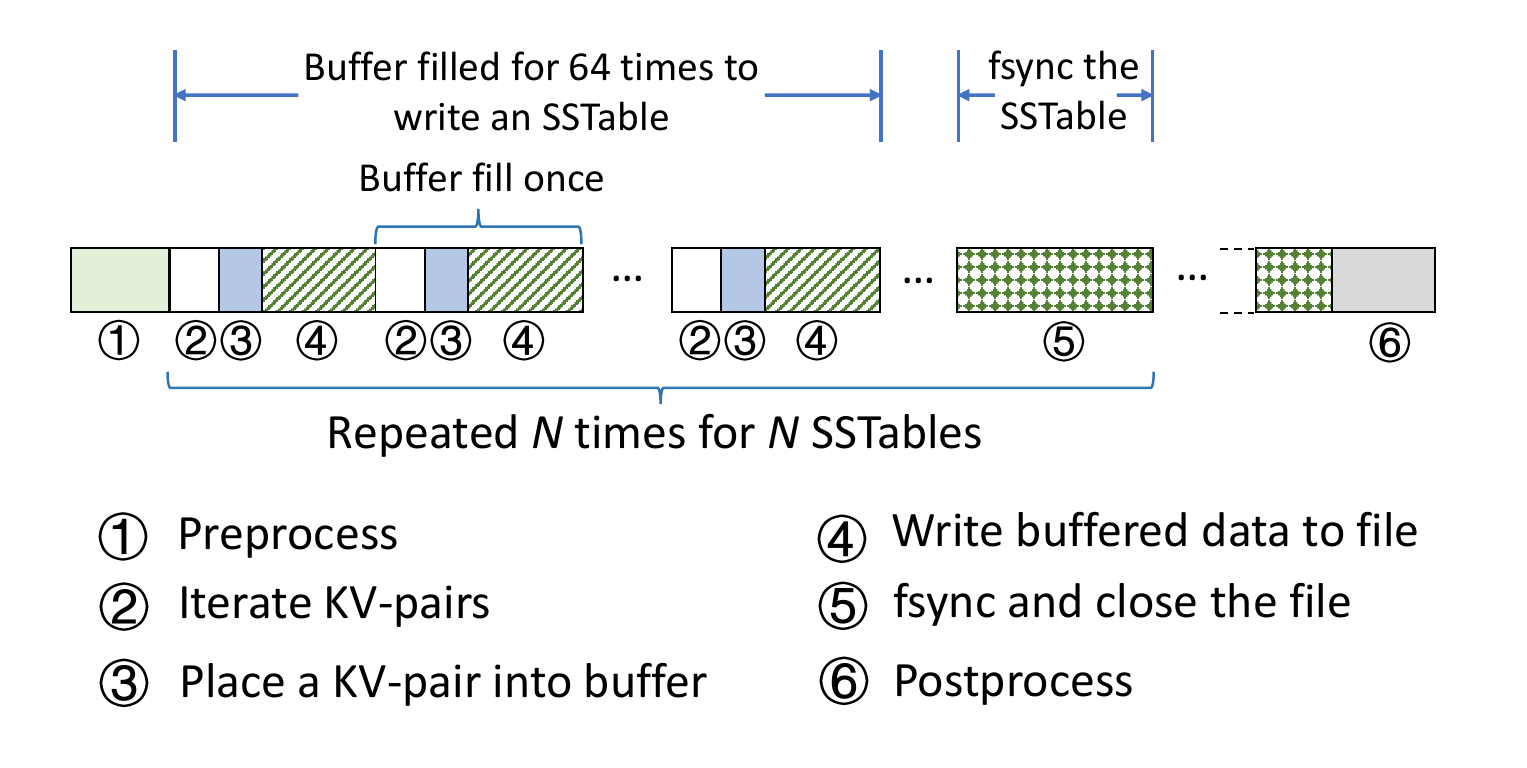}}
        \caption{The flow of a compaction generating $N$ \SSTs}
        \label{fig:Design:CompactionProcess}
    \end{center}
\end{figure}

\textbf{\textfrak{O1}:} \underline{\it Compaction severely degrades LSM-tree's performance.}

By analyzing the source code of RocksDB, we find that the system may experience stalls when a large number of \SSTs are waiting for compaction or when many memtables are pending flushes. Both scenarios involve costly {\tt fsync} operations. We conduct a test using db\_bench, which is a built-in benchmark in RocksDB. We engage four concurrent clients in continuously inserting a total of 80\ GB data of KV pairs under the fillrandom workload, with each KV pair sized at 1\,KB. We perform two experiments: (1) we explicitly disable compaction (by setting {\tt disable\_auto\_compactions = true}), and (2) we further avoid triggering any flush by configuring RocksDB to maintain an exceptionally large number of memtables. Our results show that disabling compaction increases the throughput of RocksDB by 5.1$\times$, 5.5$\times$, and 5.7$\times$ on HDD, SATA SSD, and NVMe SSD, respectively. This confirms that compaction severely degrades performance regardless of the underlying storage medium, and also aligns with observations obtained in prior works \cite{LSM:SILK:ATC-2019,LSM:NobLSM:DAC-2022}. When both compaction and flush are disabled, however, the throughput further improves by only 15.3\%, 21.9\%, and 11.7\%, respectively. Thus, the impact of flush is much less significant. We next focus on analyzing the compaction, as it is the dominant performance bottleneck.

\textbf{\textfrak{O2}: } \ul{\em In a compaction, LSM-tree stalls due to keeping the user thread waiting for the kernel thread that handles synchronous file I/Os.}

Let us overhaul the compaction procedure of RocksDB.
\autoref{fig:Design:CompactionProcess} illustrates the workflow of one compaction. By default, RocksDB employs one user thread to execute a compaction job. This thread first preprocesses the involved KV pairs by loading them from input \SSTs (\mininumbercircled{1} in
\autoref{fig:Design:CompactionProcess}). It builds an iterator over these KV pairs to locate the smallest key (\mininumbercircled{2}), places the KV pair with the smallest key into a buffer, and continues fetching one with the next smallest key. Once the buffer is filled to, say, 1MB, the user thread writes the buffered KV pairs to an output SSTable (\mininumbercircled{3}\mininumbercircled{4}). It reuses this buffer until the output file reaches a predefined size limit (e.g., 64MB), and then invokes {\tt fsync} to persist the output \SST (\mininumbercircled{5}). The steps of \mininumbercircled{2} to \mininumbercircled{5} repeat until all KV pairs are written and persisted, after which the input \SSTs are deleted (\mininumbercircled{6}).

\begin{figure}[t]
    \centering
    \begin{subfigure}[t]{0.48\columnwidth}
        \includegraphics[width=\textwidth, page=1]{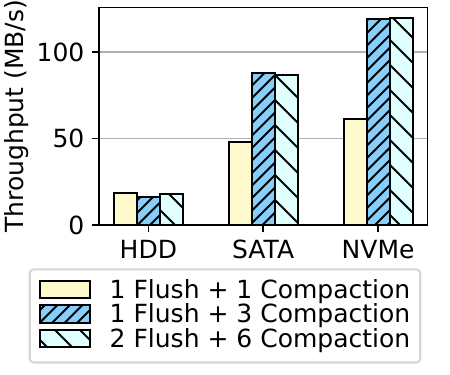}
        \caption{Varying threads for concurrent compaction jobs}
        \label{fig:Motivation:threads}
    \end{subfigure}
    \hfill
    \begin{subfigure}[t]{0.48\columnwidth}
        \includegraphics[width=\textwidth, page=1]{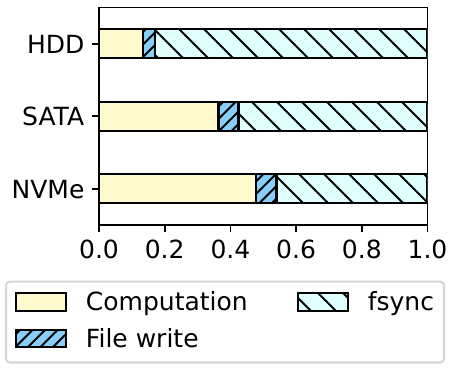}
        \caption{The breakdown of compaction with varying devices}
        \label{fig:Motivation:BreakDown}
    \end{subfigure}
    \caption{A study with RocksDB's compaction}\label{fig:mot:compaction}
\end{figure}

Note that, the steps of \mininumbercircled{2} to \mininumbercircled{5} occur sequentially and repeatedly on the critical path of compaction, whereas file writes and {\tt fsync}s are actually performed by a kernel thread. During this time, the user thread of LSM-tree remains idle, waiting for the kernel thread to complete I/Os, while the waiting time for file I/Os, particularly the {\tt fsync} operations, is non-trivial~\cite{FS:BarrierFS:FAST-2018,FS:FastCommit:ATC-2024}. We have profiled the time breakdown between user activities (\mininumbercircled{1}\mininumbercircled{2}\mininumbercircled{3}\mininumbercircled{6}) and kernel-level I/O operations (\mininumbercircled{4}\mininumbercircled{5}) for each compaction. As shown in \autoref{fig:Motivation:BreakDown}, write and {\tt fsync} I/Os take 86.5\%, 63.6\%, and 52.3\% of compaction time on average for HDD, SATA SSD, and NVMe SSD, respectively. Even with the fastest NVMe SSD, kernel thread still occupies more than half of the compaction time for I/Os, causing user thread to stall on the critical path.

\begin{figure}[t]
    \includegraphics[width=\columnwidth]{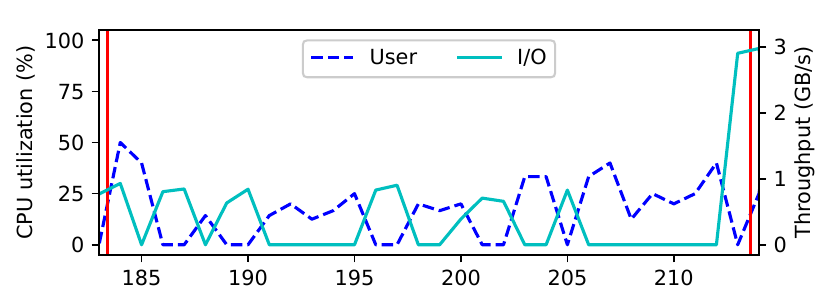}
    \caption{An illustration of CPU utilization rate of user thread against I/O throughput in a compaction job over time}\label{fig:user-io-rate}
\end{figure}

To gain a deeper understanding, we have monitored the CPU utilization of user thread and I/O bandwidth on the NVMe SSD at runtime.
\autoref{fig:user-io-rate} displays a time window of one compaction job (bounded by two vertical red bars). The dashed curve denotes CPU utilization (left Y-axis), while the solid curve indicates I/O bandwidth. High CPU utilization corresponds to sorting operations by the user thread, during which the NVMe SSD is idle. When {\tt fsync} begins, I/O bandwidth increases while CPU utilization drops, indicating that the user thread is waiting for I/O completion. These alternating up-and-down trends demonstrate that neither CPU nor SSD is fully utilized during the compaction procedure. This implies that the sequential execution model of compaction causes the under-utilization of both CPU and storage device.

By default, RocksDB employs one user thread for flush and the other one for compaction. LSM-tree-based KV stores including RocksDB generally allow users to configure the number of user threads for parallel flush and compaction jobs. RocksDB suggests that three quarters of user threads should be assigned to compaction, each handling a separate compaction job. We have increased the number of compaction threads from one to three or six. As shown in \autoref{fig:Motivation:threads}, doing so indeed improves throughput initially, but the performance soon plateaus without further increment. The reason is that, every user thread still must wait for its corresponding kernel thread to finish I/Os. Worse yet, {\tt fsync}s launched by multiple user threads serialize device-level I/Os, which instead hampers the thread-level parallelism \cite{FS:FastCommit:ATC-2024,FS:Sync+Sync:Security-2024}.

\textbf{\textfrak{O3}:} \ul{\em For data already stored in LSM-tree, existing durability ensures that it is unnecessary for the user thread to synchronously wait for the completion of {\tt fsync}s.}

Many LSM-tree-based KV stores rely on {\tt fsync}s to guarantee data durability, which is certainly essential for data integrity. However, our analysis reveals that these systems often excessively enforce durability. For LSM-tree, once a memtable is flushed, the resulting $L_0$ \SST already contains all KV pairs in a durable form. When this \SST is later involved in a compaction between $L_0$ and $L_1$, its KV pairs are re-persisted through another round of {\tt fsync} operations. In fact, every time these KV pairs participate in subsequent compactions to deeper levels $L_i$ ($i \ge 1$), they are repeatedly persisted through {\tt fsync}s once again. These redundant persisting actions introduce unnecessary I/O overhead and prolong the critical path of each compaction.

The existing durability mechanism suggests that a user thread performing compaction does not necessarily need to synchronously wait for the kernel thread to persist files through {\tt fsync}s. Instead, it can initiate the next compaction job while the persistence of files is handled asynchronously in the kernel space. There are multiple approaches to achieve this, as discussed in Section~\ref{sec:background}. To ensure both portability and efficiency across different file systems and programming environments, we will consider {io\_uring} to delegate asynchronous file write and {\tt fsync} operations to kernel threads.

While a kernel thread handles costly I/O operations in the background, the user thread proceeds with CPU-bound computations for the next compaction. This creates a parallel, pipeline-like workflow between user and kernel threads. Unlike a CPU's architectural pipeline, where the slowest stage determines the critical path, the critical path in this workflow is yet dominated by the user thread's computation, since I/Os are offloaded asynchronously to kernel threads. When multiple user threads perform concurrent compaction jobs, multiple superscalar-like workflows emerge, each consisting of one user thread paired with one kernel thread, thereby enabling higher degrees of parallelism.

\textbf{Challenges.} Although this parallelization between user and kernel threads appears promising to accelerate compaction, several challenges must be addressed. First, the separation of computations and I/Os shall not compromise the logical correctness of LSM-tree, particularly regarding data durability and accessibility. Second, as the approach relies on existing durability mechanisms, prior data copies need to be retained temporarily until their asynchronously persisted counterparts become durable; keeping these redundant copies for too long would incur excessive space overhead. As a result, we should determine a precise time at which asynchronously persisted data can be considered safely durable and old copies can be deleted. Third, when multiple accelerated compaction jobs issue {\tt fsync} requests concurrently, a burst of I/Os may congest the storage device, resulting in performance degradation. Hence, we need to incorporate an effective I/O rate control for smoother execution. Addressing these challenges constitutes the core aspects of our proposed design, namely \Pome.

\section{Design of \Pome}\label{sec:design}

\subsection{Overview}

\Pome introduces a parallel execution model to coordinate CPU computations and file I/Os for efficient storage. In a compaction, its user threads are responsible primarily for computations and no longer spin-waits for I/Os that are handled by kernel threads. By scheduling costly I/Os to be executed asynchronously,
\Pome significantly shortens the intra-compaction critical path (Section~\ref{sec:design:aio}). Therefore, the latency for serving client requests is largely reduced. At the inter-compaction level, the parallel execution model allows \Pome to overlap computations and I/Os belonging to consecutive compaction jobs (Section~\ref{sec:design:multi-overlap}). This leverages multi-core CPU and NVMe SSD to achieve high parallelism across user and kernel threads. Consequently, \Pome attains both high throughput and low latency through efficient intra- and inter-compaction parallelism.

Since \Pome completes compaction jobs at a much faster pace, it utilizes an I/O rate limiter to avoid I/O congestion (Section~\ref{sec:design:score}). Moreover,
\Pome preserves the tiered structure of LSM-tree and does not require any modification to system software or hardware. To maintain the durability and correctness of data,
\Pome tracks the generation dependency between input and output \SSTs. It performs persistence and deletion in a deferred manner, so that old \SSTs are safely removed only after their new counterparts have been fully persisted (Section~\ref{sec:design:deletion}). This transiently allows elastic capacity across involved levels but without compromising data durability or the structural rationality of LSM-tree. Overall, \Pome achieves high performance with smooth, robust execution while preserving the semantics and reliability of traditional LSM-tree.

\begin{figure}[t]
    \begin{center}
        \scalebox{1.0}{\includegraphics[width=\columnwidth, page=2]{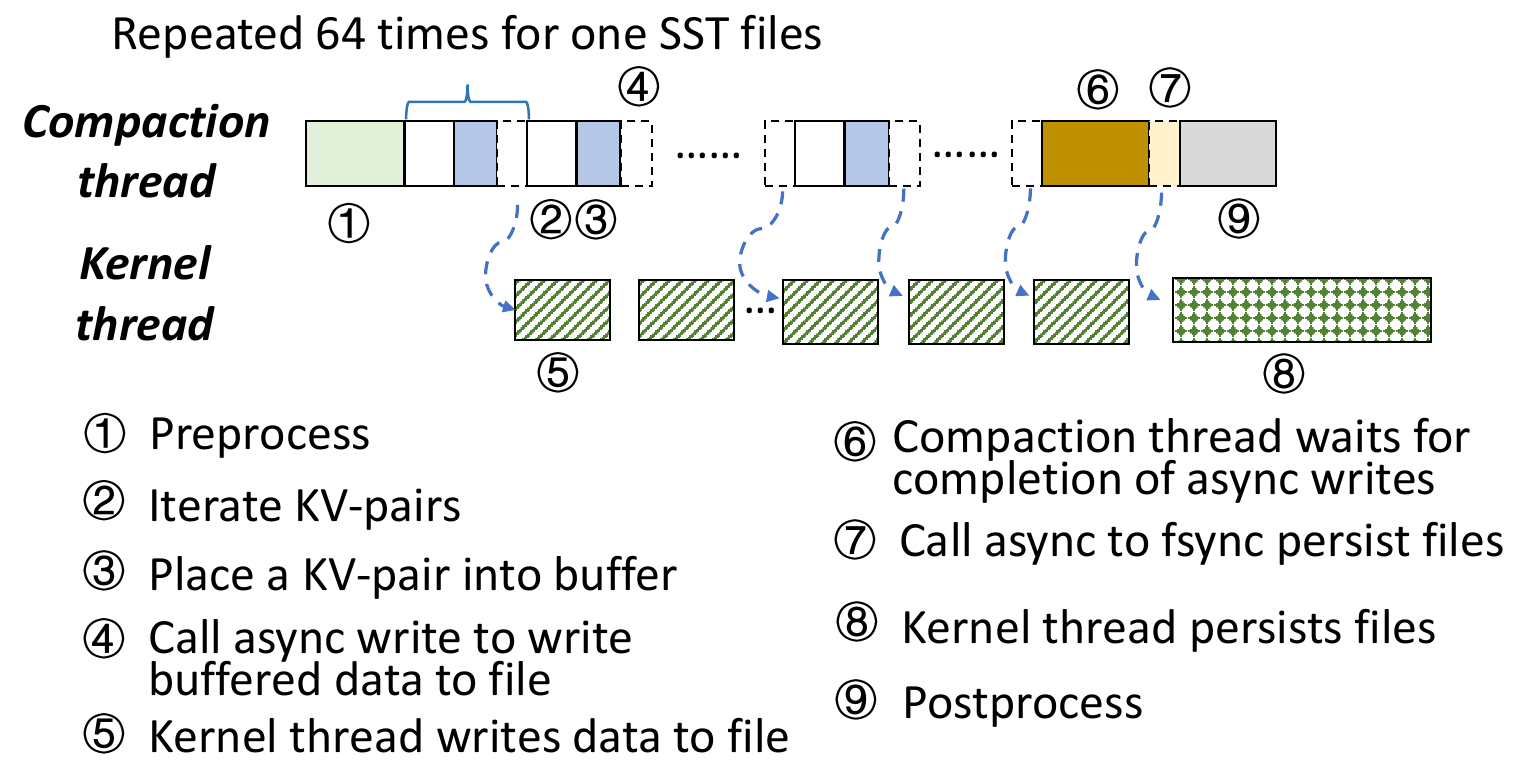}}
        \caption{The flow of \Pome's compaction}\label{fig:Design:flowchart}
    \end{center}
\end{figure}

\subsection{Intra-compaction Parallel Execution}\label{sec:design:aio}

\textbf{Accelerated compaction.} \autoref{fig:Design:flowchart} shows how \Pome handles a compaction.
\Pome chooses a victim level $L_n$ ($n \ge 0$) with a score mechanism (see Section~\ref{sec:design:score}). In brief, $L_n$ is one level that maximally exceeds its capacity limit. The user thread of \Pome preprocesses input \SSTs and loads KV pairs from them (\mininumbercircled{1} in \autoref{fig:Design:flowchart}). It then builds an iterator to scan input files and sort KV pairs in ascending order of their keys (\mininumbercircled{2}). Sorted KV pairs are placed in a buffer (\mininumbercircled{3}). When the buffer is filled up, \Pome initiates an asynchronous file write to transfer the buffered KV pairs into an output \SST (\mininumbercircled{4}). Once the \SST reaches a preset size limit, \Pome submits an asynchronous write request for the entire file and starts filling the next output \SST. A kernel thread handles write I/Os (\mininumbercircled{5}), while the user thread continues processing KV pairs without stalling.
\Pome repeats these actions until all KV pairs are submitted for asynchronous writes. It then waits for the completions of all file writes (\mininumbercircled{6}). As \Pome overlaps computations and I/Os across consecutive \SSTs, it receives timely completion signals from the underlying file system. These signals indicate that the files are accessible to user space, despite not being fully persisted in SSD yet. After all \SSTs become accessible,
\Pome launches compound asynchronous {\tt fsync}s for all of them (\mininumbercircled{7}\mininumbercircled{8}). Persisting multiple files in a batch differs from the conventional compaction that calls {\tt fsync} every time an output \SST is written. More importantly, \Pome does not synchronously wait for the completion of asynchronous {\tt fsync}s. Instead, it performs a short postprocess to conclude the compaction (\mininumbercircled{9}), such as recording the generation dependency between input and output \SSTs. As shown in \autoref{fig:Design:flowchart}, I/Os are offloaded to kernel threads, while \Pome's user threads mainly focus on computations (e.g., merge-sort). Compared with \autoref{fig:Design:CompactionProcess},
\Pome significantly shortens the critical path of compaction.

\textbf{Synchronous accessibility.}
\Pome synchronously waits for the completions of asynchronous file writes. The reason is threefold. Firstly, completed write I/Os are essential for \Pome to initiate asynchronous {\tt fsync}s, since these operations can only be issued after the corresponding file writes have finished. Secondly, one purpose of compaction for LSM-tree is to reorganize KV pairs into the sorted order across output \SSTs for efficient search and access. A completion signal returned by the file system marks all compacted KV pairs accessible, albeit without deterministic durability. Thus, the synchronous wait enforces deterministic accessibility for compacted KV pairs. Thirdly, as studied in Section~\ref{sec:motivation}, file writes take only a small portion of the total compaction time (about 6.3\%, see \autoref{fig:Motivation:BreakDown}). Since \Pome overlaps write I/Os with CPU computations (\mininumbercircled{2}\mininumbercircled{3} and \mininumbercircled{5} in \autoref{fig:Design:flowchart}), the user thread is unlikely to wait for long during this phase.

\textbf{Deferred durability.}
\Pome does not stall to wait for the completion of asynchronous {\tt fsync}s (\mininumbercircled{8} in~\autoref{fig:Design:flowchart}). It also does not immediately remove input \SSTs as conventional compaction does. Instead, \Pome retains them to preserve the durability of compacted KV pairs, since the output \SSTs are not synchronously persisted.
\Pome defers the verification of the persistence of output \SSTs until any of them is selected as an input for a future compaction. At that moment, the older input \SSTs from which those output \SSTs were generated can be safely discarded (see Section~\ref{sec:design:deletion}).

\subsection{Inter-compaction Parallel Execution}\label{sec:design:multi-overlap}

Because \Pome retains only CPU computations on the critical path of compaction, the user thread swiftly completes the current compaction and becomes ready to execute the next one shortly after. While the next compaction is performing its computations on CPU, the storage device concurrently handles {\tt fsync}s of the previous compaction. In this way, \Pome parallelizes CPU computations and file I/Os across consecutive compactions in a pipelined manner. In a conventional compaction, a user thread arranges computations and I/Os in a strictly sequential order; hence, when I/Os are being processed, the CPU core remains idle, and vice versa. By contrast, \Pome effectively engages CPU in performing computations for a new compaction while a kernel thread simultaneously handles file I/Os for the previous one. As a result, it achieves high utilization of both CPU and storage resources.

\subsection{Deferred Deletion on Asynchronous {\tt fsync}s}\label{sec:design:deletion} For a flush that transforms an immutable memtable into an $L_0$ \SST file,
\Pome synchronously invokes {\tt fsync} to persist the file onto storage. This {\tt fsync} establishes a solid foundation for the durability of KV pairs.
\Pome regards $L_0$ \SSTs as the ancestors of all subsequent \SSTs residing at lower levels, which are generated through later compactions. Each compaction can thus be viewed as a process of producing offspring output $L_{n+1}$ \SSTs from parental input $L_n$ and $L_{n+1}$ \SSTs ($n \ge 0$).

With respect to asynchronous {\tt fsync}s,
\Pome must determine an appropriate time to verify whether offspring \SST files have been fully persisted so that their parental \SSTs can be safely deleted. As the LSM-tree continuously grows with accumulating data, each \SST has a high likelihood of participating in a future compaction. Hence, \Pome decides to perform the check-up when a compaction is about to load KV pairs from its input \SSTs. It does so as follows.

Assume that a compaction takes a set of $p$ input \SSTs as parents ($p\ge1$), denoted as $\widehat{P}$ ($p\ge1$, $|\widehat{P}| = p$). All members of $\widehat{P}$ were previously generated as output \SSTs from $q$ earlier flushes or compactions ($1\le q\le p$). There is no need to examine $L_0$ \SSTs, since \Pome makes them durable synchronously. For any other file residing at $L_n$ ($n\ge1$) and participating in the current compaction,
\Pome tracks in which prior compaction, denoted as $\zeta$, the file was submitted for asynchronous {\tt fsync}. As \Pome issues asynchronous {\tt fsync}s in compound batches per compaction, it checks whether the entire batch corresponding to $\zeta$ has been persisted or not. If so, \Pome safely deletes the \SSTs that served as inputs for $\zeta$. Otherwise, \Pome synchronously waits for the completion of the asynchronous {\tt fsync}. As observed in our experiments, this is very rare in practice. Afterward, \Pome deletes the input \SSTs for $\zeta$. These input \SSTs for $\zeta$ can be viewed as the grandparents of the output \SSTs that the current compaction is about to generate.

\begin{figure}[t]
    \begin{center}
        \scalebox{1.0}{\includegraphics[width=\columnwidth]{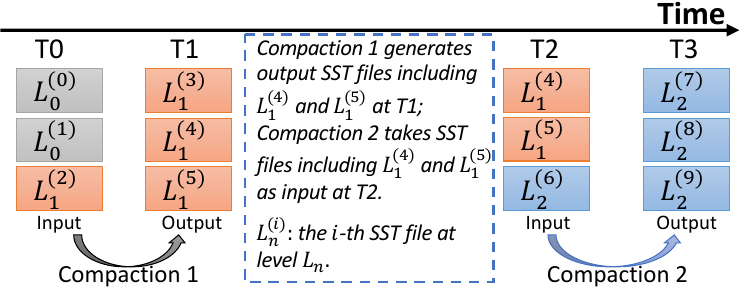}}
        \caption{An illustration of compaction flow that generates descendant SSTables}
        \label{fig:Design:compact-flow}
    \end{center}
\end{figure}

Let us use \autoref{fig:Design:compact-flow} for further illustration. It depicts two related compactions over time. At the time T$_1$, the output files $L_1^{(3)}$ of Compaction~1, together with $L_1^{(4)}$ and $L_1^{(5)}$, have not yet been made durable, so \Pome retains $L_0^{(0)}$, $L_0^{(1)}$, and $L_1^{(2)}$ until T$_2$. At T$_2$, since $L_1^{(4)}$ and $L_1^{(5)}$ participate in Compaction~2 as inputs,
\Pome checks whether the asynchronous {\tt fsync} operation issued for the three output files generated by Compaction~1 has completed. If the check returns a completion, \Pome determines that $L_1^{(3)}$, $L_1^{(4)}$, and $L_1^{(5)}$ are durable. Given their durability, the input \SST files that produced them in Compaction~1 can be safely deleted. Finally, $L_2^{(7)}$, $L_2^{(8)}$, and $L_2^{(9)}$ can be viewed as the grandchildren of $L_0^{(0)}$, $L_0^{(1)}$, and $L_1^{(2)}$.

\subsection{I/O Rate Control for Smooth Execution}\label{sec:design:score}

\Pome supports configuring multiple threads for compaction (see Section \ref{sec:impl:opt}). In this configuration, it achieves higher parallelism by employing multiple pairs of user and kernel threads to conduct concurrent computations and I/Os, respectively.

Interestingly, as \Pome substantially accelerates the compaction process, massive I/Os may accumulate at the storage device. Once the device reaches its physical I/O capability, \Pome may suffer from I/O congestion. This is common in practical with regard to excessive concurrent I/O requests \cite{10.1177/10943420231175854}. Therefore, we introduce an I/O rate limiter into \Pome to ensure smooth execution.

For efficiency, the I/O rate limiter shall be simple but effective. In developing it, we find that the effect of limiting I/O rate is equivalent to reducing the number of active compaction jobs. As mentioned, LSM-tree-based KV stores such as RocksDB select a victim level for compaction using a score mechanism~\cite{LSM:RocksDB:score}. The score of level $L_i$ ($i > 0$) is defined as the ratio of its current size to its capacity limit. The score of $L_0$ is determined as the larger value between (1) the ratio of its number of \SSTs to a predefined threshold (eight by default), and (2) the ratio of its current size to its capacity limit. RocksDB selects the level with the highest score for compaction. This score mechanism is oblivious to I/O congestion, but provides a place for us to control I/O rate. That is, if a congestion is ongoing, we can enforce a stricter selection for compaction to throttle I/O rate; if the congestion relieves, we will revoke back.

We accordingly enhance the score mechanism as follows. First, to detect if I/O congestion indeed occurs,
\Pome monitors the write throughputs of both WAL and $L_0$, which reflect client-facing service performance. Both throughputs are considered jointly for robustness. If both drop to approximately half of their upper-bound limits that have been measured under the condition of disabled compaction, \Pome claims an I/O congestion. Second, in case of I/O congestion, \Pome adjusts the calculation of every $L_i$'s score ($i > 0$) by halving it until reaching a predefined {\em lower bound}. Reduced scores decrease the likelihood of triggering compaction jobs, thereby throttling I/O activity. Meanwhile, because overlapping key ranges exist among $L_0$ \SSTs and compactions between $L_0$ and $L_1$ help to reorganize keys for faster searches~\cite{LSM:SILK:ATC-2019}, the calculation of $L_0$'s score is retained. Third, when I/O congestion alleviates as indicated by monitoring WAL and $L_0$ throughputs,
\Pome gradually restores each $L_i$'s score by doubling it. The predefined lower bound determines how aggressively \Pome restricts compaction jobs: a value that is too low or too high degrades performance. An appropriate lower bound can be empirically obtained through profiling and tuning on a specific platform.

Essentially, the I/O rate limiter curbs non-urgent compaction jobs at $L_i$ ($i > 0$) when I/O traffic becomes heavy enough to interfere with foreground client services. By incorporating this I/O rate limiter, \Pome maintains high operational efficiency, particularly when concurrent compactions are enabled.

\section{Implementation and Discussions}\label{sec:implementation}

We implement and prototype \Pome with RocksDB (Section~\ref{sec:impl:impl}). We also comprehensively consider multiple aspects to optimize and enhance it (Section~\ref{sec:impl:opt}).

\begin{figure}[t]
    \begin{center}
        \scalebox{1.0}{\includegraphics[width=\columnwidth]{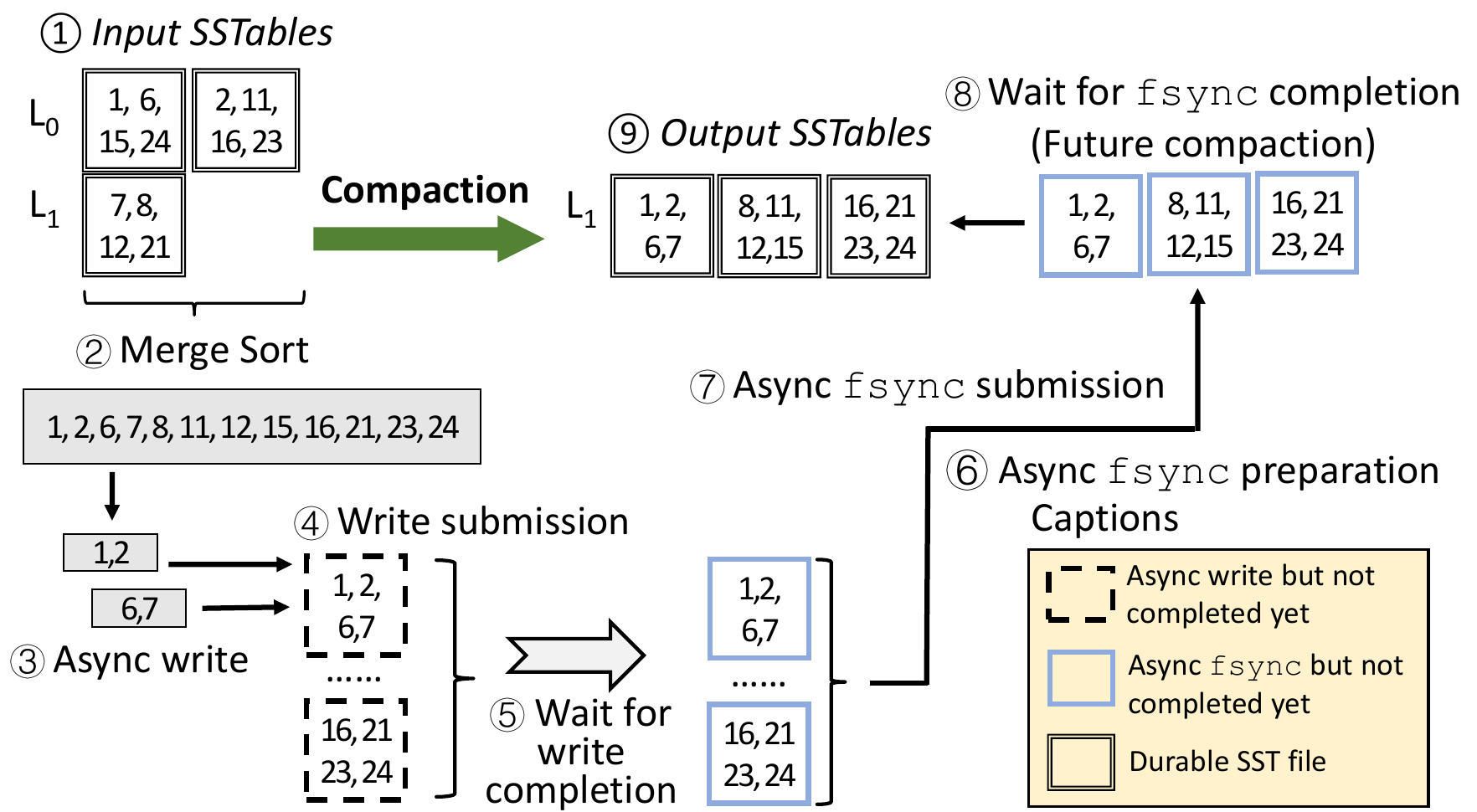}}
        \caption{An example of \Pome's compaction}\label{fig:Design:example}
    \end{center}
\end{figure}

\subsection{Implementation of \Pome}\label{sec:impl:impl}

{\bf Overview.} Doing asynchronous I/Os to revolutionize the procedure of compaction is orthogonal to other optimizations proposed to enhance LSM-tree. We take RocksDB to prototype \Pome while the ideas of \Pome can be applied to other LSM-tree variants. We mainly make use of the io\_uring to implement asynchronous writes and {\tt fsync}s for \Pome. Overall, the core functions of \Pome add or change about 1,624 lines of code (LOC) in RocksDB version 7.10.0.

{\bf Compaction procedure.}
\Pome follows RocksDB to 1) flush an immutable memtable as an $L_0$ \SST, 2) maintain background threads for flush and compaction jobs, and 3) calculate scores to choose an overfilled level and input \SSTs with key ranges overlapped for compaction. ~\autoref{fig:Design:example} illustrates eight main steps in which \Pome deals with an compaction. In particular, \Pome uses the io\_uring's structures such as {\tt uring\_queue} to collect data for each \SST file. It calls io\_uring's interfaces such as {\tt io\_uring\_prep\_fsync} and {\tt io\_uring\_submit} to prepare an asynchronous {\tt fsync} and submit an I/O request, respectively.

{\bf Deferred check-up and deletion.} At the beginning of a compaction, \Pome checks if input parental \SST files are already durable (\mininumbercircled{8}). If so, it removes grandparental \SST files by inserting them into a collection vector that RocksDB has managed for the purpose of deleting files. We note that the check-up does not cost much time. For example, in handling the aforementioned test of putting 80GB of KV pairs,
\Pome spent overall 749.1 seconds, out of which all check-up actions cost about 0.01 ms. Such a time cost is negligible.

{\bf Version and state tracking.} RocksDB has a Manifest file with an in-memory Version to record the change of \SST files (see~\autoref{fig:bg:Put}). As \Pome decouples the accessibility and durability for an $L_n$ \SST file ($n\ge1$), we modify the Manifest file such that it only updates the state of each $L_n$ \SST file when the async threads return successfully.

\subsection{Optimizations and Complements}\label{sec:impl:opt}

{\bf Concurrent compactions.} As mentioned,
\Pome maintains background threads to do flush and compaction jobs. It makes computations of current user thread execute on a CPU core while a kernel thread of io\_uring is simultaneously handling I/Os with storage for prior compaction, without blocking the user thread for compaction. Today multi-core and many-core CPUs have gathered momentum. NVMe SSD also contains numerous hardware queues for parallel I/O streams~\cite{SSD:NVMeDirect:HotStorage-2016} while Linux kernel has {\tt blk-mq} with multiple software queues~\cite{footnote:blk-mq,LSM:SpanDB:FAST-2021}.
\Pome supports the configuration of multiple threads concurrently conducting compaction jobs. Its parallel execution model, when deployed with multiple compaction threads, can exploit the parallelism capabilities of both CPU and NVMe SSD for effectual concurrent executions.

{\bf I/O polling.} NVMe SSD embraces much shorter access latency than SATA SSDs. Many researchers have used the I/O polling mechanism, instead of conventional I/O interrupts, to interact with NVMe SSD~\cite{SSD:NVMeDirect:HotStorage-2016,SSD:polling:ATC-2018,LSM:SpanDB:FAST-2021,10.1145/3731569.3764804}. In implementing \Pome, we also consider I/O polling with NVMe SSD and io\_uring.

{\bf Failed I/Os.} I/O errors might take place over time. When an I/O operation fails, the conventional synchronous I/O model helps LSM-tree handle the error in a timely fashion. As \Pome waits for the completion signals of all asynchronous file writes, any I/O error occurring at these writes can be swiftly detected and processed like with conventional LSM-tree. \Pome defers the check-up of asynchronous {\tt fsync}, so detecting and handling I/O errors for {\tt fsync} are also postponed. However, even if an I/O error happens in persisting an \SST file, the durability of KV pairs stored in the file is not impaired since \Pome has retained durable parental \SST files until the check-up. Searching KV pairs in this file is also unaffected since file system accommodates KV pairs in the OS's buffer cache or SSD's internal cache.
\Pome explicitly calls {\tt fsync} for a retry to fix the I/O error. In the worst case, it regenerates and replaces one such problematic file.

{\bf Outlier \SSTs.} In unusual cases, some \SSTs, once generated in a compaction, hardly participate in subsequent compactions. This is possible when the key ranges they cover are not frequently used, i.e., {\em outliers}.
\Pome still ensures the durability of such inactive outlier \SSTs, which is the other reason why \Pome submits one request for all \SSTs generated in a compaction to schedule compound asynchronous {\tt fsync}s. As long as any one of them is to be involved in a future compaction, \Pome checks if the asynchronous {\tt fsync} has been done for all relevant \SST files. By doing so, \Pome avoids overlooking outliers, and also helps to delete their parental \SSTs. In addition, there might be a very low likelihood that outliers form a batch and have no opportunity to be compacted again.
\Pome has tracked all \SSTs asynchronously persisted with io\_uring. It schedules a specific check-up in off-peak hours for such unusual outliers.

\begin{figure*}[t]
    \centering
    \begin{subfigure}[t]{0.245\textwidth}
        \includegraphics[width=\textwidth, page=1]{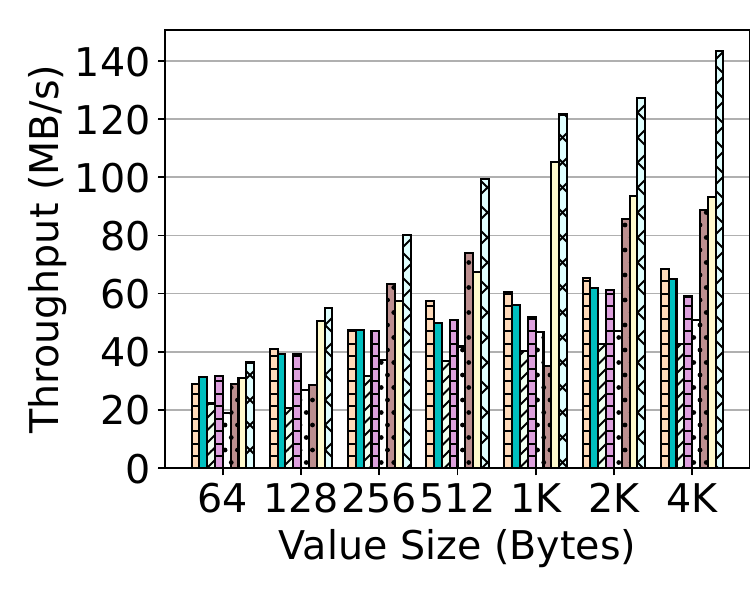}
        \caption{fillrandom}
        \label{fig:Eval:db_bench:fillrandom}
    \end{subfigure}
    \begin{subfigure}[t]{0.245\textwidth}
        \includegraphics[width=\textwidth, page=1]{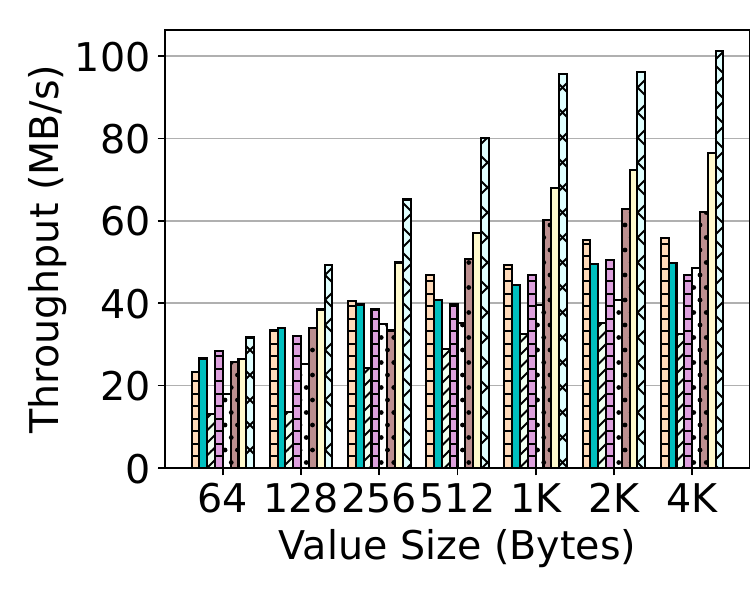}
        \caption{overwrite}
        \label{fig:Eval:db_bench:overwrite}
    \end{subfigure}
    \centering
    \begin{subfigure}[t]{0.245\textwidth}
        \includegraphics[width=\textwidth, page=1]{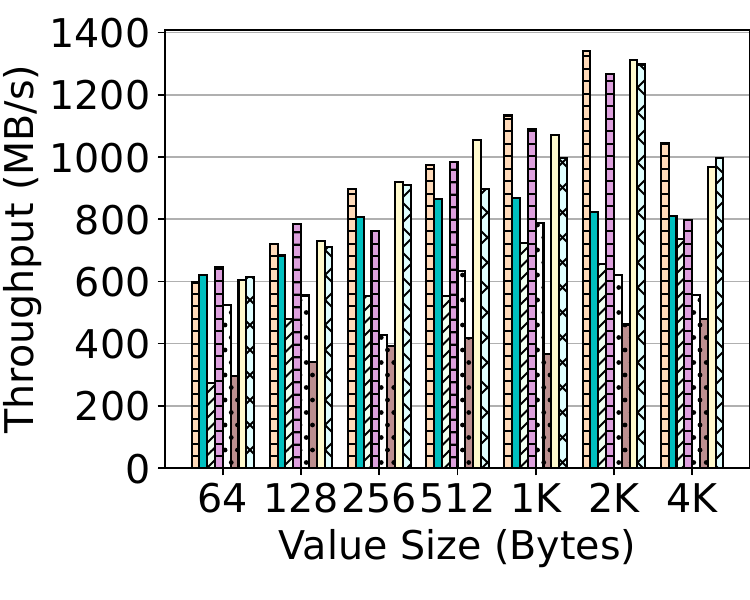}
        \caption{readseq}
        \label{fig:Eval:db_bench:readseq}
    \end{subfigure}
    \begin{subfigure}[t]{0.245\textwidth}
        \includegraphics[width=\textwidth, page=1]{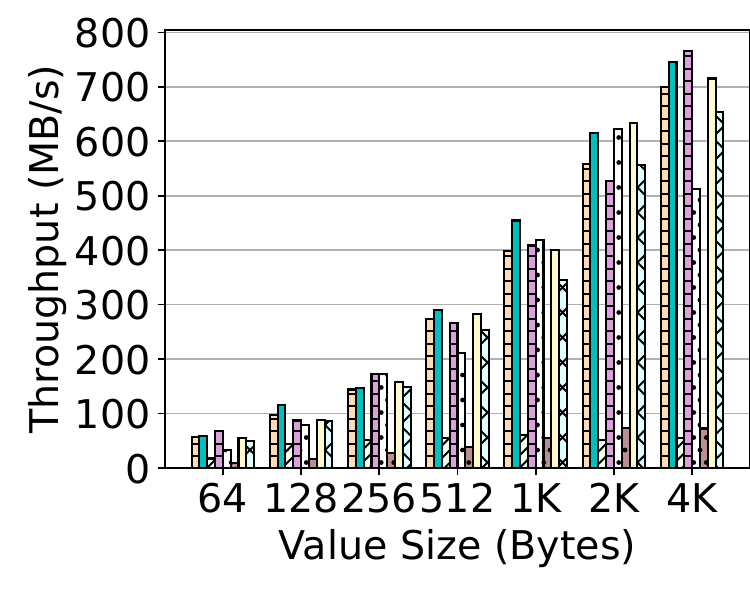}
        \caption{readrandom}
        \label{fig:Eval:db_bench:readrandom}
    \end{subfigure}
    \centering
    \begin{subfigure}[t]{1.0\textwidth}
        \includegraphics[width=\textwidth, page=1, trim=0mm 0mm 0mm 100mm, clip]{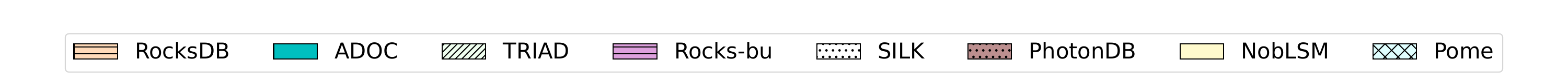}
    \end{subfigure}
    \caption{A comparison between LSM-tree variants on db\_bench's fillrandom, overwrite, readseq, and readrandom}\label{Fig:Eval:db_bench}
\end{figure*}

\section{Evaluation}\label{sec:evaluation}

We comprehensively evaluate \Pome and compare it with RocksDB as well as several SoTA LSM-tree variants that mainly adopt the sequential execution model, using both micro- and macro-benchmarks (Section \ref{subsec:evaluation_setup}). Particularly, we aim to answer following questions. \begin{itemize}	\item Does \Pome's parallel execution model  boost write performance, especially on shortening the critical path?
    Does it outperform other baselines with reshaped compaction? (Section \ref{sec:eval:write_performance})
    \item Does \Pome gain high CPU and I/O utilizations? (Section \ref{sec:eval:utilizations})
    \item Does \Pome affect data accessibility? How is its read performance? (Section \ref{sec:ReadPerformance_DataAccessibility})
    \item Does the I/O rate limiter of \Pome help to ensure high and smooth execution efficiency? (Section \ref{sec:eval:limiter})
    \item How is the performance of \Pome in processing workloads found in real-world production environments? (Section \ref{sec:eval:ycsb})
\end{itemize}

\subsection{Evaluation Setup}\label{subsec:evaluation_setup}

\textbf{Platform.} All experiments were conducted on an HP Z2 G4 workstation equipped with an Intel Core\texttrademark{} i9-9900K CPU (16 cores) and 64\,GB of DRAM. Two NVMe SSDs, i.e., Samsung 970 Pro (480\,GB) and SK Hynix PC601 (480\,GB), were used to hold WALs and other files of all key-value stores. The OS was Ubuntu~22.04.1 with Linux kernel version 6.2.7, compiled using GCC/G++~9.5.0. The version of {io\_uring} was liburing 2.3. Since XFS and {io\_uring} have been jointly optimized~\cite{uring:XFS:2}, we used XFS for all tests except for NobLSM, which must run on a customized Ext4~\cite{LSM:NobLSM:DAC-2022}. In addition, the default lower bound of \Pome's I/O rate limiter was set to $\frac{1}{8}$ based on profiling. A detailed discussion is provided in Section~\ref{sec:eval:limiter}.

\textbf{Benchmarks.} We used two benchmarks. The first one, db\_bench, is a micro-benchmark built in RocksDB that synthesizes a series of {\tt Put} and {\tt Get} requests under typical access patterns (see Section~\ref{sec:eval:micro}). The second one, YCSB~\cite{bench:YCSB}, is a macro-benchmark consisting of a suite of real-world workloads (see Section~\ref{sec:eval:ycsb}).

\textbf{Competitors.} Besides the baseline RocksDB~\cite{LSM:RocksDB}, we evaluated \Pome against several state-of-the-art LSM-tree variants that represent different research directions for improving LSM-tree performance. They are ADOC~\cite{LSM:ADOC:FAST-2023}, TRIAD~\cite{LSM:TRIAD:ATC-2017}, Rocks-bu~\cite{ioruing:rocksDB-TiKV}, SILK~\cite{LSM:SILK:ATC-2019}, PhotonDB~\cite{LSM:Alibaba:PhotonLibOS}, and NobLSM~\cite{LSM:NobLSM:DAC-2022}. All of them except NobLSM are open-source implementations based on RocksDB. NobLSM, originally built atop LevelDB, was reimplemented through modifying RocksDB and Ext4 file system to ensure a fair comparison. Below we summarize the key design characteristics of these competitors.

\emph{ADOC}: ADOC mitigates write stalls by dynamically adjusting the number of threads and SSTable sizes based on the monitored data flow within the LSM-tree. This adaptive control helps to balance processing rates and schedule background jobs more efficiently.

{\it TRIAD}: TRIAD intends to alleviate the write amplification of LSM-tree through three techniques: Firstly, it tries to separate hot KV pairs that are frequently updated from cold ones at the memtable. Secondly, it postpones a compaction until the overlap between key ranges of \SST files aggregates to some extent. Thirdly, it makes use of WAL to play the role of $L_0$ \SST file, instead of writing the same KV pairs again.

\emph{Rocks-bu}: Rocks-bu is a RocksDB variant that has used {io\_uring}. It exploits {io\_uring}'s batch I/O feature to group multiple \texttt{fsync}s, targeting the throughput improvement.

{\em SILK}: SILK focuses on I/O scheduling between insertions with memtable, flushes, and compactions, mainly for shorter tail latency. It allocates more bandwidth to internal operations, i.e., flushes and compactions, when foreground service is not heavy. It gives higher priority to flushes and compactions at lower levels (e.g., $L_0\rightarrow L_1$). Moreover, it allows compactions at lower levels to preempt ones at higher levels.

\emph{PhotonDB}: PhotonDB employs coroutines and {io\_uring} to serve one or more clients. For each client, it synchronously waits for I/O completions through {io\_uring}.

\emph{NobLSM}: NobLSM leverages the implicit asynchronous commit mechanism of Ext4 journaling to persist data, thereby removing \texttt{fsync}s from the critical path of compaction. However, it heavily depends on Ext4 journaling and requires modifications in the OS kernel (system software).

\subsection{Evaluation with Micro-benchmark}\label{sec:eval:micro} We employ db\_bench to issue four typical workloads. In particular, with each LSM-tree variant, we perform following workloads in order: fillrandom (random insertion of KV pairs), overwrite (random update of KV pairs), readseq (sequential retrieval of KV pairs), and readrandom (random retrieval of KV pairs). For each workload, we fix an overall quantity of data volume and follow the default uniform distribution to generate requests. We designate four foreground threads. Each thread puts (resp. gets) 20 GB of KV pairs for write (resp. read) requests. The choice of 80\,GB dataset is motivated by two factors. First, it generates sustained compaction activities involving significant CPU computations and file I/Os. Second, considering the write amplification effect of LSM-tree and the 480\,GB capacity of our NVMe SSDs, this data volume both stresses the memory and SSD for space consumptions and also effectively mitigates interference from internal mechanisms in the SSD, such as garbage collection and wear leveling. We set the key size as 16 B while varying the value size to be 64 B, 128 B, 256 B, 512 B, 1 KB, 2 KB, and 4 KB. By default, RocksDB employs two background threads—one for flush and one for compaction. Unless otherwise specified, all LSM-tree variants, including \Pome, use this configuration.

\subsubsection{Write Performance}\label{sec:eval:write_performance}\ \\
\indent We analyze the write performance of \Pome\ from multiple perspectives, including throughput, tail latency, and CPU-I/O utilizations, to provide an in-depth dissection of its write behavior.

\textbf{Write throughput.}
\autoref{fig:Eval:db_bench:fillrandom} and \autoref{fig:Eval:db_bench:overwrite} present the write bandwidth results. By leveraging its parallel execution model, \Pome significantly enhances write performance. For instance, across seven increasing value sizes, the throughput of \Pome is 1.3$\times$ to 2.14$\times$ higher than that of RocksDB. This improvement arises because \Pome effectively reduces the execution time of each compaction. Specifically, the parallel execution in \Pome offloads time-consuming I/O operations to kernel threads, allowing user threads to focus on computations along the critical path. This not only accelerates compaction but also increases I/O traffic utilization on the storage device. For example, with 1\,KB values under the {fillrandom} workload, RocksDB completes 2{,}414 compaction jobs, while \Pome completes 2{,}832. Moreover, the average write I/O traffic per compaction job in \Pome is 2.2$\times$ that of RocksDB.

\textbf{Varying value sizes.} With increasing value sizes,
\Pome consistently demonstrates superior performance. Given a fixed data volume, larger values result in fewer KV pairs and thus reduce CPU computations required for merge-sort. Consequently, the critical path of compaction is further shortened by \Pome, enabling it to surpass SoTA LSM-tree variants that still sequentially conduct computations and I/Os. For example, under the {fillrandom} workload with 1\,KB values, the throughput of \Pome is 2.2$\times$, 3.0$\times$, 2.3$\times$, 2.6$\times$, 2.0$\times$, 3.5$\times$, and 1.2$\times$ higher than that of RocksDB, ADOC, TRIAD, Rocks-bu, SILK, PhotonDB, and NobLSM, respectively.

\begin{table}[t]
    \centering
    \caption{The 99P latency ($\mu$s) with different value sizes}
    \resizebox{\columnwidth}{!}{
    \begin{tabular}{@{\extracolsep{\fill}}lrrrrrrrr}
        \toprule
        \small Value  &
        \multicolumn{1}{c}{\small Rocks-} &
        \multirow{2}{*}{\small ADOC} &
        \multirow{2}{*}{\small TRIAD} &
        \multicolumn{1}{c}{\small Rocks-} &
        \multirow{2}{*}{\small SILK} &
        \multicolumn{1}{c}{\small Photon-} &
        \multirow{1}{*}{\small Nob-} &
        \multirow{2}{*}{\small Pome} \\
        \small size &
        \multicolumn{1}{c}{\small DB} & ~ & ~ &
        \multicolumn{1}{c}{\small bu} & ~ &
        \multicolumn{1}{c}{\small DB} &
        \small LSM \\
        \midrule
        \small 64B  & 9.6     & 9.5     & 12.6     & 15.3    & 22.5  & 10.0    & 18.3    & 8.3   \\
        \small 256B & 20.0    & 26.3    & 32.5     & 26.4    & 12.3  & 29.5    & 24.2    & 11.6  \\
        \small 1KB  & 48.0    & 66.5    & 82.3     & 41.2    & 26.0  & 45.3    & 34.5    & 24.8  \\
        \small 4KB  & 2{,}031.2 & 2{,}315.4 & 2{,}415.6  & 1{,}167.9 & 92.3  & 1{,}956.9 & 1{,}233.4 & 61.6  \\
        \bottomrule
    \end{tabular} }
    \label{table:Eval:99p}
\end{table}

\textbf{Tail latency.} The client-facing latency is a metric that directly reflects the length of critical path. Unlike other LSM-tree variants that focus on scheduling compactions without addressing the interaction between computation and I/O, \Pome decouples and parallelizes them, achieving both higher throughput and lower latency. To examine the impact of \Pome's parallel execution model, we have recorded the 99th-percentile (99P) tail latency while each LSM-tree variant was serving db\_bench workloads.
\autoref{table:Eval:99p} shows the results with the {fillrandom} workload.

By removing synchronous costly file I/Os from the critical path, \Pome dramatically reduces the tail latency. It generally surpasses other LSM-tree variants, including SILK, which was explicitly designed to mitigate latency spikes in LSM-tree~\cite{LSM:SILK:ATC-2019}. With 1\,KB values, the 99P tail latency of \Pome is 48.3\%, 62.7\%, 69.9\%, 45.2\%, 4.5\%, 39.8\%, and 28.1\% lower than that of RocksDB, ADOC, TRIAD, Rocks-bu, SILK, PhotonDB, and NobLSM, respectively. These latency reductions are consistent with the superior throughput results of \Pome, indicating the efficacy of its parallel execution strategy.

Another observation from \autoref{table:Eval:99p} is that, with larger values, e.g., 4\,KB, the tail latency surges for all LSM-tree variants except \Pome and SILK. A similar phenomenon was also obtained in prior studies~\cite{LSM:SILK:ATC-2019}. This is explained by the working style of db\_bench, which continuously issues requests per client in the best-effort manner. Meanwhile, LSM-tree enforces a fixed capacity limit at each level. Once a level becomes full, a compaction is triggered. Given the same capacity limits for levels, larger values fill up them at a faster pace and trigger more frequent compactions, leading to more stalls and higher latency. Without specific mechanisms to reshape or reschedule compactions, such as those in \Pome and SILK, latency spikes become severer. Nonetheless, since \Pome radically reshapes the execution model at both inter- and intra-compaction levels, it achieves better latency reduction than SILK that merely reschedules at the coarser granularity of compaction job.

\begin{table}[t]
    \centering
    \caption{The 99P latency ($\mu$s) over various background jobs}
    \resizebox{\columnwidth}{!}{					\begin{tabular}{@{\extracolsep{\fill}}lrrrrrrrr}
        \toprule
        \small Background  &
        \multicolumn{1}{c}{\small Rocks-}
        & \multirow{2}{*}{\small ADOC}
        & \multirow{2}{*}{\small TRIAD}
        & \multicolumn{1}{c}{\small Rocks-}
        & \multirow{2}{*}{\small SILK}
        & \multicolumn{1}{c}{\small Photon-}
        &  \multirow{1}{*}{\small Nob-}
        &  \multirow{2}{*}{\small Pome} \\
        \small job number &\multicolumn{1}{c}{\small DB}&~&~&\multicolumn{1}{c}{\small bu}&~&\multicolumn{1}{c}{\small DB}&\small LSM\\ 				\midrule
        \small 2  & 2064.5     & 2152.2     & 2962.0     & 1234.0   & 822.9  & 2534.5    & 1835.2   & 1034.0   \\
        \small 4 & 48    & 66.5    & 82.3     & 45.3    & 26.0  & 223.7    & 34.5    & 24.8  \\
        \small 8  & 58.4    & 60.0    & 92.4     & 73.8    & 22.0  & 205.9    & 45.6    & 19.4  \\
        \small 16  & 48.3 & 50.3 & 136.1  & 88.2 & 26.2  & 85.2 & 47.3 & 22.1  \\
        \bottomrule
    \end{tabular} }
    \label{table:Eval:99p:bgjobs}
\end{table}

\textbf{Varying threads.} We next evaluate the impact of varying the number of background threads on performance. Specifically, we configure 2, 4, 8, and 16 threads for concurrent flush and compaction jobs, using the {fillrandom} workload with 1\,KB values.
\autoref{table:Eval:99p:bgjobs} shows the 99P tail latencies for all configurations. With its stable parallel execution model and I/O rate limiter, \Pome consistently maintains low tail latency, without the latency spikes observed in PhotonDB or TRIAD. For example, with eight background threads, the tail latencies of RocksDB and other baselines are 3.0$\times$, 3.1$\times$, 4.8$\times$, 3.8$\times$, 1.1$\times$, 10.6$\times$, and 2.4$\times$ higher than that of \Pome, respectively.

\begin{figure}[t]
    \centering
    \includegraphics[width=\columnwidth]{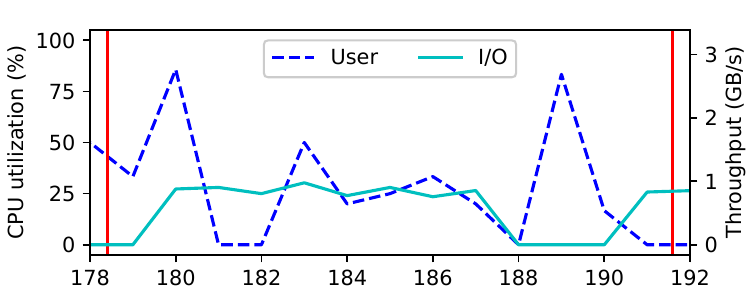}
    \caption{CPU utilization of user threads versus I/O throughput in a compaction job over time.}
    \label{fig:Eval:CPU-IO}
\end{figure}

\textbf{CPU and I/O utilizations.}\label{sec:eval:utilizations} One of the primary goals of parallelizing computation and I/O in \Pome is to keep both CPU and storage device fully utilized. To complement \autoref{fig:user-io-rate}, we present in \autoref{fig:Eval:CPU-IO} the CPU and I/O utilizations monitored during a representative compaction job executed by \Pome with 1\,KB values under the {fillrandom} workload. In the diagram, the dashed line denotes the CPU utilization rate measured in percentage of the user thread (left Y-axis), while the solid line indicates the I/O bandwidth measured in GB/s (right Y-axis).

We get three key observations from two curves shown in \autoref{fig:Eval:CPU-IO}. First, the compaction of \Pome completes in a shorter duration compared to RocksDB (see \autoref{fig:user-io-rate}). Second, by avoiding synchronously waiting for \texttt{fsync}s, \Pome achieves higher CPU utilization for the user thread. Third, \Pome sustains significantly greater I/O bandwidth because kernel thread continuously processes asynchronous I/O requests. These results demonstrate that \Pome effectively overlaps CPU computations and file I/Os, fully exploiting system resources to raise the overall efficiency for LSM-tree.

\subsubsection{Read Performance}\label{sec:ReadPerformance_DataAccessibility}\ \\
\indent In this section, we present a comprehensive evaluation of \Pome under read-intensive workloads. We first validate the crash consistency of \Pome, followed by its data accessibility verification and read performance analysis.

\textbf{Crash consistency.}\label{subsec:CrashConsistencyEvaluation} To verify the crash consistency of \Pome, we refer to prior works \cite{LSM:NobLSM:DAC-2022} and issue a sudden power-off using the Linux command \texttt{halt -f -p -n} while executing the {fillrandom} workload in db\_bench. The test is repeated five times for both RocksDB and \Pome. In all cases, the KV pairs persisted in the WAL or flushed to \SSTs are recoverable and retrievable in both systems.

By default, neither RocksDB nor \Pome enforces synchronous \texttt{fsync}s on WALs. The key difference lies in how \Pome manages compaction outputs: it does not wait for the durability of new \SST files before proceeding, nor does it immediately delete input \SST files. Instead, \Pome performs a post-check before deleting them, ensuring that every KV pair moving from $L_n$ to $L_{n+1}$ ($n \ge 0$) is traceable and durable through its generation dependency tracking mechanism. This design guarantees crash recoverability while preserving compaction efficiency.

\textbf{Data accessibility.}\label{subsec:eval:access} To test if \Pome maintains full data accessibility despite it decoupling the immediate accessibility from delayed persistence, we conduct read experiments using the {readrandom} workload. A client first inserts 20\,GB of KV pairs with varying value sizes using the {fillrandom} workload, followed by random key searches (i.e., readrandom). Note that newer versions of RocksDB (since version 6.2) no longer guarantee that db\_bench must search for existing keys and each time it may change a seed for generating random keys. We hence fix the seed to ensure reproducible and identical search sequences across different LSM-tree variants.

As told by
\autoref{table:Eval:keys-found}, \Pome retrieves the same number of KV pairs as RocksDB with all four value sizes, confirming that they gain equivalent data accessibility. During each compaction, \Pome waits for the completion of all asynchronous file writes before marking output \SSTs visible. Even though some data may still reside in the OS's buffer cache or the SSD's internal cache, the completion signals from the file system ensure that all files are visible (accessible) to both the OS and LSM-tree.

\begin{table}[t]
    \centering
    \caption{Accessibility test with the {readrandom} workload.}
    \resizebox{1.02\columnwidth}{!}{		\begin{tabular}{lrrrrrr}
        \toprule
        \multicolumn{1}{c}{Value} & \multicolumn{3}{c}{RocksDB} & \multicolumn{3}{c}{\Pome}\\  \cline{2-7}
        \multicolumn{1}{c}{size} & No. of KV & No. of KV & \multirow{2}{*}{Rate} & No. of KV & No. of KV & \multirow{2}{*}{Rate} \\
        & pairs put & pairs got & & pairs put & pairs got & \\
        \midrule
        64B  & 268,435,456 & 169,678,058 & 63.2\% & 268,435,456 & 169,678,058 & 63.2\% \\
        256B & 78,951,604  & 49,910,913  & 63.2\% & 78,951,604  & 49,910,913  & 63.2\% \\
        1KB  & 20,648,881  & 13,055,759  & 63.2\% & 20,648,881  & 13,055,759  & 63.2\% \\
        4KB  & 5,222,479   & 3,298,559   & 63.2\% & 5,222,479   & 3,298,559   & 63.2\% \\
        \bottomrule
    \end{tabular}\label{table:Eval:keys-found} }
\end{table}

\textbf{Read performance.}\label{subsec:eval:ReadPerformanceEvaluation} We next evaluate \Pome using two read-intensive workloads, i.e., readseq and {readrandom}. As shown in \autoref{fig:Eval:db_bench:readseq} and \autoref{fig:Eval:db_bench:readrandom}, the read throughput of \Pome is overall comparable to that of RocksDB, with only marginal degradation in certain cases, which we plan to optimize in the future using techniques like KV caching \cite{LSM:AC-key-cache:FAST-2020}.

\begin{figure}[t]
    \centering
    \includegraphics[width = \columnwidth]{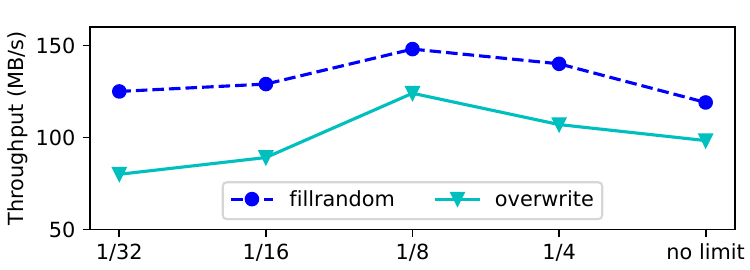}
    \caption{Impact of the I/O rate limiter on \Pome.}
    \label{fig:Eval:rate}
\end{figure}

\subsubsection{The impact of I/O rate limiter}\label{sec:eval:limiter}\ \\
\indent We also verify how the I/O rate limiter enhances \Pome. The lower bound is varied among {\em no limit}, $\frac{1}{4}$, $\frac{1}{8}$ (default), $\frac{1}{16}$, and $\frac{1}{32}$, representing a spectrum from relaxed to stringent rate limiting.
\autoref{fig:Eval:rate} displays the throughput curves of \Pome under the {fillrandom} and {overwrite} workloads, where the value size is 1\,KB. The rightmost points for {\em no limit} (without the I/O rate limiter) illustrate that as more compactions generate increasing I/O traffic, the throughput of \Pome declines due to the lack of I/O control, since excessive I/Os collectively congest the SSD and in turn degrade the front-end service to clients. By throttling I/Os to a reasonable extent, specifically, using a $\frac{1}{8}$ lower bound on our platform,
\Pome achieves the highest performance on both workloads. A balanced I/O rate control prevents both excessive SSTable accumulation and overly frequent compactions, thereby ensuring sustained throughput and stable client-facing performance.

\subsection{Macro-benchmark Evaluation}\label{sec:eval:ycsb}

The YCSB benchmark suite includes six core workloads reflecting production scenarios, i.e., A (50\%/50\% read/write), B (95\%/5\% read/write), C (100\% read), D (95\%/5\% read/insert), E (95\%/5\% range query/insert), and F (50\%/50\% read-modify-write/read). These workloads span from write-dominant to read-dominant patterns, with the default KV size being 1 KB (10 fields per value, 100 bytes per field). Following prior studies~\cite{LSM:PebblesDB:SOSP-2017,LSM:NobLSM:DAC-2022}, we execute the workloads in the following order: Load-A, A, B, C, F, D, Load-E, and E. Load-A and Load-E initialize the system by clearing existing data and inserting 50 million KV pairs (1 KB each, ~50 GB total). Each subsequent workload executes 10 million requests.

\begin{figure}[t]
    \centering
    \begin{subfigure}[t]{0.45\textwidth}
        \includegraphics[width=\textwidth,page=1]{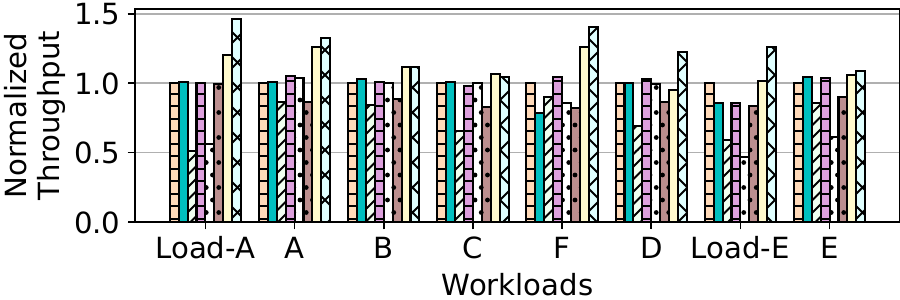}
        \includegraphics[width=\textwidth, page=1, trim=0mm 0mm 0mm 0mm, clip]{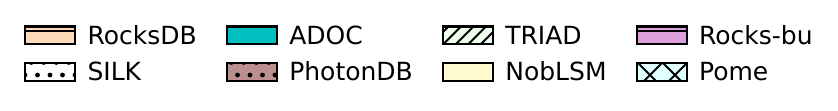}

        \label{fig:Eval:YCSB:LOAD-A}
    \end{subfigure}
    \caption{Normalized throughput comparison of LSM-tree variants under YCSB workloads.}
    \label{fig:Eval:YCSB}
\end{figure}

Because of the substantial variations between YCSB workloads, we take the result of RocksDB as the unit for normalization.
\autoref{fig:Eval:YCSB} presents the normalized throughput results of all LSM-tree variants. \Pome consistently achieves higher or comparable performance across different workloads, demonstrating its effectiveness under diverse access patterns.

\autoref{fig:Eval:YCSB} further shows two observations. First, under write-dominant workloads such as Load-A and Load-E, \Pome again surpasses competing LSM-tree variants. For example, with Load-A, \Pome achieves 1.5$\times$, 1.4$\times$, 2.8$\times$, 1.5$\times$, 2.6$\times$, 1.5$\times$, and 1.2$\times$ higher throughput than RocksDB, ADOC, TRIAD, Rocks-bu, SILK, PhotonDB, and NobLSM, respectively. This performance advantage is attributed to \Pome's parallelized compactions and asynchronous I/O handling. Second, in mixed read/write workloads such as A and F, \Pome maintains strong performance, achieving 32.6\% and 40.7\% higher throughput than RocksDB, respectively. Therefore, \Pome is also performant in a workload mixed of write and read requests.

\begin{figure}[t]

    \begin{subfigure}[t]{0.485\columnwidth}
        \includegraphics[width=\textwidth, page=1]{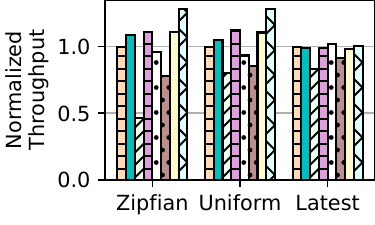}
        \caption{Worklaod A}
        \label{fig:Eval:A-Distribution-Sensitivity}
    \end{subfigure}
    \begin{subfigure}[t]{0.485\columnwidth}
        \includegraphics[width=\textwidth, page=1]{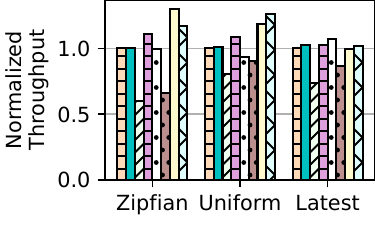}
        \caption{Worklaod F}
        \label{fig:Eval:F-Distribution-Sensitivity}
    \end{subfigure}
    \begin{subfigure}[t]{0.485\textwidth}
        \includegraphics[width=\textwidth, page=1, trim=0mm 0mm 0mm 0mm, clip]{./figs_for_tpds/pdfs/YCSB/YCSB-legend.pdf}
    \end{subfigure}
    \caption{A comparison between LSM-tree variants with different key distributions}\label{fig:Eval:Distribution-Sensitivity}
    \vspace{-3ex}
\end{figure}

Next, we evaluate \Pome and other LSM-tree variants under different key distributions. Besides the default uniform distribution, we consider other two common ones, i.e., Zipfian and the latest. Without loss of generality, we present the results with A and F workloads of YCSB. As mentioned, both workloads consist of a 50\%/50\% mix of write and read requests. The throughput presented in~\autoref{fig:Eval:Distribution-Sensitivity} show that \Pome consistently outperforms other LSM-tree variants under both Zipfian and uniform distributions. Under the latest distribution, however, \Pome exhibits performance comparable to that of RocksDB. This observation arises because the latest distribution continually selects the most recently accessed data for operations. Consequently, most write and read requests are served directly from the memtables and block cache of LSM-tree. Due to frequent updates targeting the same keys at the foreground memtable, compactions are triggered less frequently. As a result, the advantages brought by \Pome's accelerated compaction mechanism are diminished.
\section{Related Works} \label{sec:relatedWork}

We have discussed several representative LSM-tree variants in Section~\ref{sec:evaluation}. Beyond those, researchers have explored a range of approaches to optimize compaction, reduce stalls, and improve I/O efficiency for LSM-tree. For example, Zhang et al.~\cite{LSM:PipelinedCompaction:IPDPS-2014} dissected the internals of compaction and analyzed it at the block granularity (4~KB). They pipelined computation and synchronous I/Os for consecutive blocks within each \SST. However, their approach required multiple storage devices for sufficient bandwidth and demanded modifications across user and kernel spaces for storage management. Moreover, the pipeline lacked stability—small data segments in one or a few blocks are difficult for the CPU and storage device to process at a steady rate. In contrast, \Pome operates at the \SST\ granularity (in megabytes), requires no changes to the storage stack, and provides better viability and stability.

Key-value separation effectively reduces compaction frequency. For instance, WiscKey stores only pointers to actual values in \SST files to avoid rewriting large values during compaction~\cite{LSM:WiscKey:FAST-2016}. Concurrent compaction techniques have also been proposed. For example, p$^2$KVS~\cite{LSM:p2KVS:EuroSys-2022} partitions the KV space and manages multiple LSM-tree instances that execute compaction jobs concurrently.
\Pome's new execution model complements such techniques, and they can be jointly applied to achieve even higher performance.

Some studies investigated the impact of {\tt fsync} on LSM-tree performance~\cite{LSM:BoLT:Middleware-2020,LSM:NobLSM:DAC-2022}. BoLT aggregates multiple {\tt fsync} operations into a single large one to reduce overhead~\cite{LSM:BoLT:Middleware-2020}, but this large {\tt fsync} still lies on the critical path. As mentioned, NobLSM~\cite{LSM:NobLSM:DAC-2022} leverages Ext4’s implicit commit to replace synchronous {\tt fsync}s and uses journaling to track durability. It is tightly coupled with Ext4 and requires handcrafted kernel modifications. It also suffers from I/O congestion. In contrast, \Pome introduces a parallel execution model without any dependency on specific system software, and incorporates an adaptive I/O rate limiter to ensure robustness. In addition, regarding the usefulness of asynchronous I/O, researchers have utilized it in many domains. For example, Malkums et al.~\cite{10.1145/3731545.3735116} used io\_uring to optimize the checkpoint restoration for HPC applications.

Researchers have also explored other dimensions to optimize LSM-tree. Some studies accelerate the computations of compaction using GPUs~\cite{GPComp:Algorithm:TPDS-2025}. Others tailor LSM-tree designs for byte-addressable non-volatile memory (NVM)~\cite{TriangleKV:LDS:TPDS-2022} or shingled magnetic recording (SMR) devices~\cite{SEALDB:LDS:TPDS-2019}. LDS~\cite{LDS:FS:TPDS-2022} further integrates LSM-tree management directly with raw block devices, bypassing the traditional file system to eliminate index interference and fully exploit sequential I/O potential. Recent studies such as EcoTune~\cite{EcoTune:LSM:SIGMOD-2025} have revealed that compaction strategies influence both read and write performance, motivating coordinated optimization across compaction scheduling and I/O control. This insight aligns with the design philosophy of \Pome, where we employ a dynamic I/O rate limiter to adaptively regulate the number of active compaction jobs and preserve throughput stability.

\section{Conclusion}\label{sec:conclusion}

The sequential execution model between computations and I/Os is widely adopted in data storage systems. In this paper, we particularly study LSM-tree-based KV stores. Compaction is essential and important for LSM-tree. The sequential execution of each compaction incurs a long critical path and severely impairs the performance of LSM-tree. We accordingly develop \Pome with a new parallel execution model. In brief,
\Pome employs a user thread for computations and schedules I/Os to be asynchronously conducted by a kernel thread. It hence parallelizes computations and I/Os for consecutive compactions, and shortens the critical path of serving client requests. This also improves the utilizations of CPU and storage device as \Pome keeps both busy over time. We prototype \Pome and thoroughly evaluate it. Extensive experiments show that
\Pome significantly boosts the performance of RocksDB and outperforms state-of-the-art LSM-tree variants.

\begin{acks}
	We sincerely thank the reviewers and TPC of the 35th ACM International Symposium on High-Performance Parallel and Distributed Computing (HPDC 2026) for their valuable comments and suggestions.
	We also express sincere gratitude to Mr. Qun Xu and Mr. Tianming Wen of the School of Information Science and Technology, ShanghaiTech University for hardware support.
	This work was jointly supported by
	Natural Science Foundation of Shanghai under Grants No. 23ZR1442300 and ShanghaiTech Startup Funding.
\end{acks}

\bibliographystyle{ACM-Reference-Format}
\bibliography{main}

@misc{AnIntrod6:online,
author = {Bijan Mottahedeh},
title = {An Introduction to the io\_uring Asynchronous {I/O} Framework},
howpublished = {\url{https://blogs.oracle.com/linux/post/an-introduction-to-the-io-uring-asynchronous-io-framework}},
month = {May},
year = {2020},
day = {27},
}

@article{10.1177/10943420231175854,
	author = {Dongarra, Jack and Tourancheau, Bernard and Jeannot, Emmanuel and Pallez, Guillaume and Vidal, Nicolas},
	title = {{IO}-aware Job-Scheduling: Exploiting the Impacts of Workload Characterizations to select the Mapping Strategy},
	year = {2023},
	issue_date = {Jul 2023},
	publisher = {Sage Publications, Inc.},
	address = {USA},
	volume = {37},
	number = {3–4},
	issn = {1094-3420},
	url = {https://doi.org/10.1177/10943420231175854},
	doi = {10.1177/10943420231175854},
	journal = {Int. J. High Perform. Comput. Appl.},
	month = jul,
	pages = {213–228},
	numpages = {16}
}

@inproceedings {IO:XRP:OSDI-2022,
	author = {Yuhong Zhong and Haoyu Li and Yu Jian Wu and Ioannis Zarkadas and Jeffrey Tao and Evan Mesterhazy and Michael Makris and Junfeng Yang and Amy Tai and Ryan Stutsman and Asaf Cidon},
	title = {{XRP}: {In-Kernel} Storage Functions with {eBPF}},
	booktitle = {16th USENIX Symposium on Operating Systems Design and Implementation (OSDI 22)},
	year = {2022},
	isbn = {978-1-939133-28-1},
	address = {Carlsbad, CA},
	pages = {375--393},
	publisher = {USENIX Association},
	month = {July},
}

@inproceedings {LSM:ELECT:FAST-2024,
	author = {Yanjing Ren and Yuanming Ren and Xiaolu Li and Yuchong Hu and Jingwei Li and Patrick P. C. Lee},
	title = {{ELECT}: Enabling Erasure Coding Tiering for {LSM-tree-based} Storage},
	booktitle = {22nd USENIX Conference on File and Storage Technologies (FAST 24)},
	year = {2024},
	isbn = {978-1-939133-38-0},
	address = {Santa Clara, CA},
	pages = {293--310},
	publisher = {USENIX Association},
	month = feb,
}

@inproceedings{10.1145/3588195.3595949,
	author = {Saha, Manoj P. and Desai, Omkar and Kim, Bryan S. and Bhimani, Janki},
	title = {Leveraging Keys In Key-Value {SSD} for Production Workloads},
	year = {2023},
	isbn = {9798400701559},
	publisher = {Association for Computing Machinery},
	address = {New York, NY, USA},
	url = {https://doi.org/10.1145/3588195.3595949},
	doi = {10.1145/3588195.3595949},
	booktitle = {Proceedings of the 32nd International Symposium on High-Performance Parallel and Distributed Computing},
	pages = {327–328},
	numpages = {2},
	location = {Orlando, FL, USA},
	series = {HPDC '23}
}

@inproceedings{10.1145/3731569.3764804,
	author = {Zhou, Yanbo and Xu, Erci and Su, Anisa and Harris, Jim and Manzanares, Adam and Swanson, Steven},
	title = {Sleeping with One Eye Open: Fast, Sustainable Storage with {Sandman}},
	year = {2025},
	isbn = {9798400718700},
	publisher = {Association for Computing Machinery},
	address = {New York, NY, USA},
	url = {https://doi.org/10.1145/3731569.3764804},
	doi = {10.1145/3731569.3764804},
	booktitle = {Proceedings of the ACM SIGOPS 31st Symposium on Operating Systems Principles},
	pages = {496–511},
	numpages = {16},
	location = {Lotte Hotel World, Seoul, Republic of Korea},
	series = {SOSP '25}
}

@inproceedings{10.1145/3731545.3735116,
	author = {Malkmus, Zackary and Tan, Nigel and Lumsden, Ian and Assogba, Kevin and Rafique, M. Mustafa and Nicolae, Bogdan and Michela Taufer, Michela},
	title = {On Optimizing Checkpoint Restoration for {HPC} Applications: Leveraging Merkle Trees and Asynchronous {I/O}},
	year = {2025},
	isbn = {9798400718694},
	publisher = {Association for Computing Machinery},
	address = {New York, NY, USA},
	url = {https://doi.org/10.1145/3731545.3735116},
	doi = {10.1145/3731545.3735116},
	booktitle = {Proceedings of the 34th International Symposium on High-Performance Parallel and Distributed Computing},
	articleno = {27},
	numpages = {2},
	location = {University of Notre Dame Conference Facilities, Notre Dame, IN, USA},
	series = {HPDC '25}
}

@inproceedings {FS:Sync+Sync:Security-2024,
author = {Qisheng Jiang and Chundong Wang},
title = {{Sync+Sync}: A Covert Channel Built on fsync with Storage},
booktitle = {33rd USENIX Security Symposium (USENIX Security 24)},
year = {2024},
isbn = {978-1-939133-44-1},
address = {Philadelphia, PA},
pages = {3349--3366},
publisher = {USENIX Association},
month = aug
}

@inproceedings {FS:FastCommit:ATC-2024,
	author = {Harshad Shirwadkar and Saurabh Kadekodi and Theodore Tso},
	title = {{FastCommit}: resource-efficient, performant and cost-effective file system journaling},
	booktitle = {2024 USENIX Annual Technical Conference (USENIX ATC 24)},
	year = {2024},
	isbn = {978-1-939133-41-0},
	address = {Santa Clara, CA},
	pages = {157--171},
	publisher = {USENIX Association},
	month = jul
}

@inproceedings{FS:BarrierFS:FAST-2018,
	author = {Won, Youjip and Jung, Jaemin and Choi, Gyeongyeol and Oh, Joontaek and Son, Seongbae and Hwang, Jooyoung and Cho, Sangyeun},
	title = {Barrier-Enabled {IO} Stack for Flash Storage},
	year = {2018},
	isbn = {9781931971423},
	publisher = {USENIX Association},
	address = {USA},
	booktitle = {Proceedings of the 16th USENIX Conference on File and Storage Technologies},
	pages = {211-226},
	numpages = {16},
	location = {Oakland, CA, USA},
	series = {FAST'18},
}

@misc{Asio-lib,
	author = {Christopher Kohlhoff},
	title = {{ASIO} {C++} Library},
	howpublished = {\href{https://think-async.com/Asio/}{https://think-async.com/Asio/}},
	month = {July},
	year = {2024},
	day = {7},
}

@misc{uring:XFS:2,
	author = { Michael Larabel},
	title = {Linux 5.20 To Support Async Buffered Writes For {XFS} + IO\_uring For Big Performance Boost},
	howpublished = {\href{https://www.phoronix.com/news/Linux-520-XFS-uring-Async-Buff}{https://www.phoronix.com/news/Linux-520-XFS-uring-Async-Buff}},
	month = {June },
	year = {2022},
	day = {22},
}

@inproceedings {SSD:polling:ATC-2018,
	author = {Bo Peng and Haozhong Zhang and Jianguo Yao and Yaozu Dong and Yu Xu and Haibing Guan},
	title = {{MDev-NVMe}: A {NVMe} Storage Virtualization Solution with Mediated {Pass-Through}},
	booktitle = {2018 USENIX Annual Technical Conference (USENIX ATC 18)},
	year = {2018},
	isbn = {978-1-939133-01-4},
	address = {Boston, MA},
	pages = {665--676},
	publisher = {USENIX Association},
	month = jul,
}

@inproceedings{SSD:NVMeDirect:HotStorage-2016,
	author = {Kim, Hyeong-Jun and Lee, Young-Sik and Kim, Jin-Soo},
	title = {{NVMeDirect}: A User-Space {I/O} Framework for Application-Specific Optimization on {NVMe} {SSDs}},
	year = {2016},
	publisher = {USENIX Association},
	address = {USA},
	booktitle = {Proceedings of the 8th USENIX Conference on Hot Topics in Storage and File Systems},
	pages = {41-45},
	numpages = {5},
	location = {Denver, CO},
	series = {HotStorage'16},
}

@MISC{footnote:blk-mq,
	title        = "Multi-Queue Block {IO} Queueing Mechanism ({\tt blk-mq})",
	booktitle    = "The Linux Kernel",
	author       = "{The kernel development community}",
	howpublished = "\url{https://www.kernel.org/doc/html/latest/block/blk-mq.html}",
	language     = "en",
}

@misc{ioruing:rocksDB-TiKV,
	author = {PingCAP-Hackthon2019-Team17},
	title = {{IO}-uring speed the {RocksDB} \& {TiKV}},
	note = {\href{https://openinx.github.io/ppt/io-uring.pdf}{https://openinx.github.io/ppt/io-uring.pdf}},
	year = {2019},
	month = {October},
}

@misc{aio:linux-man,
	author = { 	Linux man page },
	title = {aio - {POSIX} asynchronous {I/O} overview},
	note = {\href{https://linux.die.net/man/7/aio}{https://linux.die.net/man/7/aio}},
	month = {July},
	day = {11},
	year = {2024},
}

@misc{iouring,
	author = {Jens Axboe},
	title = {Efficient {IO} with io\_uring},
	note = {\href{https://kernel.dk/io_uring.pdf}{https://kernel.dk/io\_uring.pdf}},
	month = {October},
	day  = {25},
	year = {2019},

}

@inproceedings {LSM:SpanDB:FAST-2021,
	author = {Hao Chen and Chaoyi Ruan and Cheng Li and Xiaosong Ma and Yinlong Xu},
	title = {{SpanDB}: A Fast, {Cost-Effective} {LSM-tree} Based {KV} Store on Hybrid Storage},
	booktitle = {19th USENIX Conference on File and Storage Technologies (FAST 21)},
	year = {2021},
	isbn = {978-1-939133-20-5},
	pages = {17--32},
	publisher = {USENIX Association},
	month = feb,
}

@inproceedings {LSM:AC-key-cache:FAST-2020,
	author = {Fenggang Wu and Ming-Hong Yang and Baoquan Zhang and David H.C. Du},
	title = {{AC-Key}: Adaptive Caching for {LSM}-based Key-Value Stores},
	booktitle = {2020 {USENIX} Annual Technical Conference ({USENIX} {ATC} 20)},
	year = {2020},
	isbn = {978-1-939133-14-4},
	pages = {603--615},
	publisher = {{USENIX} Association},
	month = {July},
}

@inproceedings{LSM:NobLSM:DAC-2022,
author = {Dang, Haoran and Ye, Chongnan and Hu, Yanpeng and Wang, Chundong},
	title = {{NobLSM}: An {LSM}-Tree with Non-Blocking Writes for {SSDs}},
	year = {2022},
	isbn = {9781450391429},
	publisher = {Association for Computing Machinery},
	address = {New York, NY, USA},
	doi = {10.1145/3489517.3530470},
	booktitle = {Proceedings of the 59th ACM/IEEE Design Automation Conference (DAC  22)},
	pages = {403-408},
	numpages = {6},
	location = {San Francisco, California},

}

@inproceedings{LSM:PebblesDB:SOSP-2017,
	author = {Raju, Pandian and Kadekodi, Rohan and Chidambaram, Vijay and Abraham, Ittai},
	title = {{PebblesDB}: Building Key-Value Stores Using Fragmented Log-Structured Merge Trees},
	year = {2017},
	isbn = {9781450350853},
	publisher = {Association for Computing Machinery},
	address = {New York, NY, USA},

	doi = {10.1145/3132747.3132765},
	booktitle = {Proceedings of the 26th Symposium on Operating Systems Principles},
	pages = {497-514},
	numpages = {18},
	location = {Shanghai, China},
	series = {SOSP '17},
}

@inproceedings {LSM:TRIAD:ATC-2017,
	author = {Oana Balmau and Diego Didona and Rachid Guerraoui and Willy Zwaenepoel and Huapeng Yuan and Aashray Arora and Karan Gupta and Pavan Konka},
	title = {{TRIAD}: Creating Synergies Between Memory, Disk and Log in Log Structured Key-Value Stores},
	booktitle = {2017 {USENIX} Annual Technical Conference ({USENIX} {ATC} 17)},
	year = {2017},
	isbn = {978-1-931971-38-6},
	address = {Santa Clara, CA},
	pages = {363--375},

	publisher = {{USENIX} Association},
	month = {July},
}

@misc{LSM:LevelDB,
	title={{LevelDB}},
	author={Sanjay Ghemawat and Jeff Dean},
	year={2011},
	month= {March},
	day = {19},
	note={\href{https://github.com/google/leveldb}{https://github.com/google/leveldb}}
}

@misc{LSM:RocksDB,
	title={{RocksDB}},
	author={{Facebook Database Engineering Team}},
	year={2017},
	month= {October},
	day = {18},
	note={\href{https://rocksdb.org/}{https://rocksdb.org/}}
}

@misc{LSM:RocksDB:score,
	title={{RocksDB's leveled compaction}},
	author={{Facebook Database Engineering Team}},
	year={2023},
	month= {November},
	day = {14},
	note={\href{https://github.com/facebook/rocksdb/wiki/Leveled-Compaction}{https://github.com/facebook/rocksdb/wiki/Leveled-Compaction}}
}

@inproceedings{bench:YCSB,
	author = {Cooper, Brian F. and Silberstein, Adam and Tam, Erwin and Ramakrishnan, Raghu and Sears, Russell},
	title = {Benchmarking Cloud Serving Systems with {YCSB}},
	booktitle = {Proceedings of the 1st ACM Symposium on Cloud Computing},
	series = {SoCC '10},
	year = {2010},
	isbn = {978-1-4503-0036-0},
	location = {Indianapolis, Indiana, USA},
	pages = {143--154},
	numpages = {12},
	acmid = {1807152},
	publisher = {ACM},
	address = {New York, NY, USA},
}

@inproceedings {LSM:in-SSD-index:OSDI-2021,
	author = {Jinhyung Koo and Junsu Im and Jooyoung Song and Juhyung Park and Eunji Lee and Bryan S. Kim and Sungjin Lee},
	title = {Modernizing File System through In-Storage Indexing},
	booktitle = {15th {USENIX} Symposium on Operating Systems Design and Implementation ({OSDI} 21)},
	year = {2021},
	isbn = {978-1-939133-22-9},
	pages = {75--92},

	publisher = {{USENIX} Association},
	month = {July},
}

@inproceedings {LSM:FPGA-accelerated-compaction:FAST-2020,
	author = {Teng Zhang and Jianying Wang and Xuntao Cheng and Hao Xu and Nanlong Yu and Gui Huang and Tieying Zhang and Dengcheng He and Feifei Li and Wei Cao and Zhongdong Huang and Jianling Sun},
	title = {{FPGA}-Accelerated Compactions for {LSM}-based Key-Value Store},
	booktitle = {18th {USENIX} Conference on File and Storage Technologies ({FAST} 20)},
	year = {2020},
	isbn = {978-1-939133-12-0},
	address = {Santa Clara, CA},
	pages = {225--237},

	publisher = {{USENIX} Association},
	month = {February},
}

@inproceedings{LSM:BoLT:Middleware-2020,
	author = {Kim, Dongui and Park, Chanyeol and Lee, Sang-Won and Nam, Beomseok},
	title = {{BoLT}: Barrier-Optimized {LSM}-Tree},
	year = {2020},
	isbn = {9781450381536},
	publisher = {Association for Computing Machinery},
	address = {New York, NY, USA},
	doi = {10.1145/3423211.3425676},
	booktitle = {Proceedings of the 21st International Middleware Conference (Middleware '20)},
	pages = {119-133},
	numpages = {15},
	location = {Delft, Netherlands},
}

@inproceedings {LSM:WiscKey:FAST-2016,
	author = {Lanyue Lu and Thanumalayan Sankaranarayana Pillai and Andrea C. Arpaci-Dusseau and Remzi H. Arpaci-Dusseau},
	title = {{WiscKey}: Separating Keys from Values in {SSD}-conscious Storage},
	booktitle = {14th {USENIX} Conference on File and Storage Technologies ({FAST} 16)},
	year = {2016},
	isbn = {978-1-931971-28-7},
	address = {Santa Clara, CA},
	pages = {133-148},
	month = {February},
}

@ARTICLE{SEALDB:LDS:TPDS-2019,
	author={Yao, Ting and Tan, Zhihu and Wan, Jiguang and Huang, Ping and Zhang, Yiwen and Xie, Changsheng and He, Xubin},
	journal={IEEE Transactions on Parallel and Distributed Systems},
	title={{SEALDB}: An Efficient {LSM}-tree Based KV Store on {SMR} Drives with Sets and Dynamic Bands},
	year={2019},
	volume={30},
	number={11},
	pages={2595-2607},
	doi={10.1109/TPDS.2019.2918219}}

@ARTICLE{TriangleKV:LDS:TPDS-2022,
	author={Ding, Chen and Yao, Ting and Jiang, Hong and Cui, Qiu and Tang, Liu and Zhang, Yiwen and Wan, Jiguang and Tan, Zhihu},
	journal={IEEE Transactions on Parallel and Distributed Systems},
	title={{TriangleKV}: Reducing Write Stalls and Write Amplification in {LSM}-Tree Based KV Stores With Triangle Container in {NVM}},
	year={2022},
	volume={33},
	number={12},
	pages={4339-4352},
	doi={10.1109/TPDS.2022.3188268}}

@ARTICLE{GPComp:Algorithm:TPDS-2025,
	author={Zhou, Hao and Chen, Yuanhui and Zeng, Wu and Cui, Lixiao and Wang, Gang and Liu, Xiaoguang},
	journal={IEEE Transactions on Parallel and Distributed Systems},
	title={{GPComp}: Using {GPU} and {SSD-GPU} Peer to Peer {DMA} to Accelerate {LSM}-Tree Compaction for Key-Value Store},
	year={2025},
	volume={36},
	number={9},
	pages={1920-1936},
	doi={10.1109/TPDS.2025.3586616}}

@misc{LSM:Alibaba:PhotonLibOS,
title = {Photon{LibOS}},
author = {{Alibaba Cloud}},
note = {\url{https://github.com/alibaba/PhotonLibOS}},
year = {2022},
month = {July},
day = {27},
}

@misc{photonDB:coroutine,
author = {Bob Chen},
title = {200 lines of code to rewrite the 600,000 lines {RocksDB} into a coroutine program},
year = {2022},
month = {December},
note = {\href{https://github.com/facebook/rocksdb/issues/11017}{https://github.com/facebook/rocksdb/issues/11017}},
day = {5},
}

@inproceedings{LSM:p2KVS:EuroSys-2022,
author = {Lu, Ziyi and Cao, Qiang and Jiang, Hong and Wang, Shucheng and Dong, Yuanyuan},
title = {{p$^2$KVS}: A Portable 2-Dimensional Parallelizing Framework to Improve Scalability of Key-Value Stores on {SSDs}},
year = {2022},
isbn = {9781450391627},
publisher = {Association for Computing Machinery},
address = {New York, NY, USA},
booktitle = {Proceedings of the Seventeenth European Conference on Computer Systems},
pages = {575-591},
numpages = {17},
location = {Rennes, France},
series = {EuroSys '22}
}

@inproceedings {LSM:ADOC:FAST-2023,
author = {Jinghuan Yu and Sam H. Noh and Young{\textendash}ri Choi and Chun Jason Xue},
title = {{ADOC}: Automatically Harmonizing Dataflow Between Components in  Log-Structured Key-Value  Stores for Improved Performance},
booktitle = {21st USENIX Conference on File and Storage Technologies (FAST 23)},
year = {2023},
isbn = {978-1-939133-32-8},
address = {Santa Clara, CA},
pages = {65--80},
publisher = {USENIX Association},
month = feb,
}

@inproceedings {LSM:SILK:ATC-2019,
author = {Oana Balmau and Florin Dinu and Willy Zwaenepoel and Karan Gupta and Ravishankar Chandhiramoorthi and Diego Didona},
title = {{SILK}: Preventing Latency Spikes in {Log-Structured} Merge {Key-Value} Stores},
booktitle = {2019 USENIX Annual Technical Conference (USENIX ATC 19)},
year = {2019},
isbn = {978-1-939133-03-8},
address = {Renton, WA},
pages = {753--766},
publisher = {USENIX Association},
month = jul,
}

@INPROCEEDINGS{LSM:PipelinedCompaction:IPDPS-2014,
  author={Zhang, Zigang and Yue, Yinliang and He, Bingsheng and Xiong, Jin and Chen, Mingyu and Zhang, Lixin and Sun, Ninghui},
  booktitle={2014 IEEE 28th International Parallel and Distributed Processing Symposium (IPDPS 14)},
  title={Pipelined Compaction for the {LSM}-Tree},
  year={2014},
  volume={},
  number={},
  pages={777-786},
  publisher = {IEEE Press},
  doi={10.1109/IPDPS.2014.85},
}

@article{EcoTune:LSM:SIGMOD-2025,
	author = {Wang, Hengrui and Qiu, Jiansheng and Yuan, Fangzhou and Zhang, Huanchen},
	title = {Rethinking The Compaction Policies in {LSM}-trees},
	year = {2025},
	issue_date = {June 2025},
	publisher = {Association for Computing Machinery},
	address = {New York, NY, USA},
	volume = {3},
	number = {3},
	doi = {10.1145/3725344},
	journal = {Proc. ACM Manag. Data},
	month = jun,
	articleno = {207},
	numpages = {26}
}

@article{LDS:FS:TPDS-2022,
author = {Mei, Fei and Cao, Qiang and Jiang, Hong and Tian, Lei},
title = {{LSM}-Tree Managed Storage for Large-Scale Key-Value Store},
journal={IEEE Transactions on Parallel and Distributed Systems},
year={2019},
volume={30},
number={2},
pages={400-414},
doi={10.1109/TPDS.2018.2864209},
}

\appendix
\section{Appendix for Artifact Evaluation} \label{sec:appendix:repro} This appendix provides detailed instructions for reproducing the experimental results reported in this paper, including the hardware platform, software environment, benchmark workloads, baselines, the source code links, and compilation steps.

\subsection{Hardware Platform} As mentioned in Section \ref{sec:evaluation}, all experiments are conducted on a machine with the following configuration: \begin{itemize}
    \item \textbf{Machine:} HP Z2 G4 workstation;
    \item \textbf{CPU:} Intel Core{\texttrademark} i9-9900K (16 cores);
    \item \textbf{Main Memory:} 64GB DRAM;
    \item \textbf{Storage Devices:}
    \begin{itemize}
        \item Samsung 970 Pro NVMe SSD (480GB);
        \item SK Hynix PC601 NVMe SSD (480GB).
    \end{itemize}
\end{itemize} Two NVMe SSDs are used to store write-ahead logs (WALs) and all other persistent files for the evaluated key-value stores.

\subsection{Software Environment} The software stack used in our experiments is summarized below: \begin{itemize}
    \item \textbf{Operating System:} Ubuntu 22.04.1;
    \item \textbf{Linux Kernel:} 6.2.7;
    \item \textbf{Compiler:} GCC/G++ 9.5.0;
    \item \textbf{io\_uring Library:} liburing 2.3;
    \item \textbf{File Systems:}
    \begin{itemize}
        \item XFS (default);
        \item Ext4 (only for NobLSM).
    \end{itemize}
\end{itemize} Because XFS and io\_uring have been jointly optimized~\cite{uring:XFS:2}, we have used XFS is used for all LSM-tree variants except NobLSM. NobLSM requires a customized Ext4 file system and handcrafted kernel modifications~\cite{LSM:NobLSM:DAC-2022}.

\subsection{Source Code and Compilation} Here we introduce how \Pome can be compiled. The source code of \Pome is available at:

\vspace{2ex}
\noindent\framebox[\columnwidth][s]{
\parbox{\dimexpr\linewidth-2\fboxsep-2\fboxrule}{
\textcolor{blue}{\ \ \ \ \ \ \ \ \ \ \ \ \ \ \ \ \ \ \url{https://github.com/toast-lab/LSM-Pome}}. }}
\vspace{0.5ex}

\subsubsection*{Environment Requirements} \begin{itemize}
    \item \textbf{Linux kernel}: With regard to the rapid evolution of \texttt{io\_uring}, we recommend Linux kernel version 6.2 or higher. At minimum, Linux kernel version 5.2 is required for functional correctness.
    \item \textbf{Dependency}: This project depends on the {liburing} library, which can be obtained from: \url{https://github.com/axboe/liburing}
\end{itemize}
\begin{quote}
    \textit{Note:} Before installing {liburing}, we recommend removing any existing versions of the library from the system to avoid unpredictable runtime behavior.
\end{quote}

\subsubsection*{Compilation Steps} \begin{enumerate}
    \item Install all required dependencies and clone the project repository from the aforementioned link.
    \item Compile the project using the following commands:
    \begin{lstlisting}
    mkdir -p build && cd build
    cmake -DCMAKE_BUILD_TYPE=Release .. -DWITH_SNAPPY=1
    cmake --build .
    \end{lstlisting}
\end{enumerate}

\subsection{System Configuration} For \Pome, the default lower bound of the I/O rate limiter is set to $\frac{1}{8}$, determined empirically via profiling. A discussion of this parameter is provided in Section~\ref{sec:eval:limiter}.

\subsection{Benchmarks} \label{sec:appendix:benchmarks} Our evaluation employs both micro- and macro-benchmarks. Detailed usage instructions can be found in the respective documentation provided by their authors (developers). \begin{itemize}
    \item \textbf{db\_bench:} This micro-benchmark is built into RocksDB and synthesizes sequences of \texttt{Put} and \texttt{Get} operations under typical access patterns. No separate download is required.
    \item \textbf{YCSB:} We use the C++ implementation of the Yahoo! Cloud Serving Benchmark (YCSB)~\cite{bench:YCSB}, which can be found at \url{https://github.com/ls4154/YCSB-cpp}.
\end{itemize}

\subsection{Baseline Systems} \label{sec:appendix:baselines} We compare \Pome against RocksDB and several state-of-the-art LSM-tree variants. For reproducibility, we provide the corresponding source code links. You can compile them following the instructions provided by their authors. \begin{itemize}
    \item \textbf{RocksDB:} \cite{LSM:RocksDB}
    \url{https://github.com/facebook/rocksdb.git}
    \item \textbf{ADOC}~\cite{LSM:ADOC:FAST-2023}:
    \url{https://github.com/supermt/FEAT\_7.11.git}
    \item \textbf{TRIAD}~\cite{LSM:TRIAD:ATC-2017}:
    \url{https://github.com/epfl-labos/TRIAD.git}
    \item \textbf{Rocks-bu}~\cite{ioruing:rocksDB-TiKV}:
    \url{https://github.com/PingCAP-Hackthon2019-Team17/rocksdb.git}
    \item \textbf{SILK}~\cite{LSM:SILK:ATC-2019}:
    \url{https://github.com/theoanab/SILK-USENIXATC2019.git}
    \item \textbf{PhotonDB}~\cite{photonDB:coroutine}:
    \url{https://github.com/alibaba/PhotonLibOS.git}
    \item \textbf{NobLSM}~\cite{LSM:NobLSM:DAC-2022}:
    Source code is not publicly available; please contact the authors for access.
\end{itemize}

\end{document}